\documentclass[a4paper,fleqn,usenatbib]{mnras}

\usepackage[T1]{fontenc}
\usepackage{ae,aecompl}

\usepackage{graphicx}	% Including figure files
\usepackage{amsmath}	% Advanced maths commands
\usepackage{amssymb}	% Extra maths symbols
\usepackage{epsfig}
\usepackage[usenames]{color}
\usepackage{pdfpages}
\usepackage{array}
\usepackage{multirow}
\newcolumntype{P}[1]{>{\centering\arraybackslash}p{#1}}
\newcolumntype{L}[1]{>{\arraybackslash}l{#1}}
\newcolumntype{M}[1]{>{\centering\arraybackslash}m{#1}}
\usepackage{tabularx}
\usepackage{enumitem}
\newlist{tabitemize}{itemize}{1}
\setlist[tabitemize]{nosep,
                  topsep= 0pt,
                  partopsep=0pt,
                  leftmargin= *,
                  label=\textbullet,
                  before=\vspace{-0.6\baselineskip},
                  after=\vspace{-\baselineskip}
                  }
\usepackage{booktabs}
\usepackage{adjustbox}

\title[Uncertainties in the HFF]{Systematic vs. Statistical Uncertainties in Masses and Magnifications of the Hubble Frontier Fields}

\author[Raney et al.]{
Catie A. Raney,$^{1}$\thanks{E-mail: raney@physics.rutgers.edu}
Charles R. Keeton,$^{1}$
Sean Brennan,$^{1}$
and Hsin Fan${^2}$
\\
$^{1}$Department of Physics and Astronomy, Rutgers University, 136 Frelinghuysen Road, Piscataway, NJ 08854;\\
$^{2}$Department of Physics and Astronomy, Stony Brook University, Stony Brook, NY 11794\\
}

\date{Accepted XXX. Received YYY; in original form ZZZ}

\pubyear{2020}

\begin{document}
\label{firstpage}
\pagerange{\pageref{firstpage}--\pageref{lastpage}}
\maketitle

\begin{abstract}
The Hubble Frontier Fields data, along with multiple data sets obtained by other telescopes, have provided some of the most extensive constraints on cluster lenses to date. Multiple lens modeling teams analyzed the fields and made public a number of deliverables. By comparing these results, we can then undertake a unique and vital test of the state of cluster lens modeling. Specifically, we see how well the different teams can reproduce similar magnifications and mass profiles. We find that the circularly averaged mass profiles of the fields are remarkably constrained (scatter $<5\%$) at distances of 1 arcmin from the cluster core, yet magnifications can vary significantly. Averaged across the six fields, we find a bias of -6\% (-17\%) and a scatter of $\sim$40\% ($\sim$65\%) at a modest magnification of 3 (10). Statistical errors reported by individual teams are often significantly smaller than the differences among all the teams, indicating the importance of continued systematics studies in cluster lensing. 
\end{abstract}


%%%%%%%%%%%%%%%%%%%%%%%%%%%%%%%%%%%%%%%%%%%%%%%%%%

%%%%%%%%%%%%%%%%% BODY OF PAPER %%%%%%%%%%%%%%%%%%

\section{Introduction} \label{sec:intro}

Galaxy clusters are the largest gravitationally bound objects in our Universe, with masses of $10^{14}-10^{15} M_{\odot}$. They are dominated by dark matter, but are also made up of both hot gas in the intracluster medium (ICM) and hundreds to thousands of galaxies. These structures are built up by mergers of groups and other clusters of galaxies, which can give them complicated mass distributions. However, they can be very informative to study. For example, how common these extreme systems are and how mass is distributed within them can give constraints on dark matter properties. An example of the latter is the well known Bullet cluster \citep{clowe2006}, and a similar analysis has been applied to many systems since \citep[e.g.][]{bradac2008,merten2011,harvey2015}. 

Gravitational lensing can be a useful tool in studying the mass of these galaxy clusters \citep[see review by][]{hoekstra2013}. Lensing occurs when light from a background source is bent by intervening mass. Since galaxy clusters are both very massive and large on the sky, they offer a wide area over which this lensing can be detected. In the weak lensing regime, the image of the background galaxy is only very slightly stretched tangentially around the cluster. While this stretch usually cannot be seen by eye, it can be detected through statistical studies of thousands of galaxies. This allows for the mass distribution of the cluster to be constrained out to large radii, but with low resolution \citep[see e.g.][]{umetsu2014,bartelmann2017,murata2019}.

Strong lensing occurs closer to the core of the cluster, where the density is highest. In this case, the light from a background galaxy is more strongly affected, and two or more images of the galaxy are produced. These multiple images can be used to constrain the mass of the cluster within the strong lensing region, i.e. where the multiple images are found. This offers higher resolution than weak lensing, but is limited in radius \citep[][etc.]{jauzac2018,cibirka2018,andrade2019}.

In the case of strong lensing, galaxy clusters can also be used as cosmic telescopes \citep[see review by][]{kneib2011}. The multiple images produced often have a magnification that makes the images of the source appear brighter than they would without the lensing effect. Further, they can be stretched out into long arcs; this allows the study of the galaxy at a higher resolution than it would have otherwise, down to sub-kiloparsec scale \citep[e.g.][]{livermore2012,johnson2017,dunham2019}. This has been particularly useful in the study of intermediate- and high-redshift galaxies (z>6), which are intrinsically small and very faint \citep[e.g.][]{Zheng2012,coe2013,salmon2018}.

The goal of the Hubble Frontier Fields program \citep[HFF;][]{lotz2017} was to use galaxy clusters in this way to study galaxies from the first billion years of cosmic history. The program included an extensive observing campaign to produce very deep, multi-band images of six known lensing clusters. In addition, a number of other campaigns utilized different ground-based telescopes, which provided both spectroscopic and photometric data in different bands and over a wider area. Combining these produced a wealth of information on galaxies, both in the cluster and along the line of sight, as well as on candidate lensed images. The program proved successful, with a number of images found at high redshift, allowing the luminosity functions at $z\sim6$ and greater to be better estimated \citep{mcleod2016,bouwens2017,oesch2018}. 

An important part of the program was that multiple teams were invited and/or funded to model the fields. In order to determine the intrinsic properties of a lensed galaxy, e.g. its size and luminosity, one must use a lens model to determine how much it is being magnified. To do that, a model of the mass in the field must be constructed. Of course, with such complicated systems, there are many possible sources of error in the models. Some of these errors have been studied \citep[e.g.][]{host2012,johnson2016,acebron2017,chirivi2017,raney2019}, but not all of them. If many teams model the fields, some of these errors will be marginalized over, or at least explored, when combining results. 

The Hubble Frontier Fields dataset then is extremely useful, not just in creating detailed models of the fields in question, but also in comparing results from multiple teams. \citet{priewe2017} examined magnifications within the core of two HFF clusters, Abell 2744 and MACS J0416, finding high dispersion (30\% at low magnifications) between the version 3 models analyzed. \citet{remolina2018} also considered models of the field MACS J0416, though they studied scatter in RMS of images and how well old models could predict the positions of new images. \citet{meneghetti2017} generated two mock clusters, aiming to produce mass distributions that were similar to clusters of the HFF sample, in both mass and complexity. They then asked teams to model the two fields and compared the results with a variety of metrics. In the case of mock clusters, the true mass distribution is known, as are the magnifications of the images, which makes comparing the models easier than with real clusters where it is not known. However, mock clusters might not capture the full complexity of a real mass distribution.

In this work, we aim to expand on previous studies by comparing the publicly available results\footnote{https://archive.stsci.edu/prepds/frontier/lensmodels/} in the latest round (verison 4) of modeling all six HFF clusters. In particular, we examine mass profiles and magnifications. By surveying how well the models of various teams agree, we can both test the current state of the field and use the results as a way to inform future cluster lensing work. This is especially useful since it is not a given that cluster lensing studies in the future will have the amount of modeling effort that the HFF project did: only one or two teams might model a field, and thus would likely not be able to capture the full errors in magnification.

We begin this paper with an overview of the HFF modeling process in Section 2. From there, we look at mass profiles in Section 3, as well as give a brief introduction to each field. In Section 4 we examine the magnification maps submitted. We discuss results from both mass and magnification comparisons in Section 5. We conclude our findings and offer broader implications of the work in Section 6. 

\section{HFF Modeling Overview}
\subsection{Data and Process} 
In this work, we compare models created in the latest (version 4) round of modeling. The process started with teams numerically ranking candidate lensed images based on spectroscopic data, matching colors and morphologies, and whether or not a team would use an image as a constraint on their model. The images were then given a medal ranking. \textsc{Gold} images were those for which the majority of teams were confident the image was part of a lensed family and it had a spectroscopic redshift; \textsc{silver} images also required high confidence, but did not have secure spectroscopic redshifts. More tenuous images were given the \textsc{bronze} ranking, while some images received no ranking if, for example, they were added late in the process and thus not all teams ranked them. Tables of images we used to constrain our models, as well as the catalogs used for cluster member and line-of-sight galaxy selection, can be found in \citet{raney2019}. 

In creating the models, teams were left to choose their own inputs and modeling methodology. Techniques for lens modeling fall within two categories: parametric and free-form (sometimes called nonparametric). Parametric models consist of small-scale halos for galaxies and large-scale halos for dark matter and ICM/hot gas. Mass is usually assigned to galaxies using scaling relations tied to some reference galaxy, e.g. the BCG or an $L$* galaxy at the cluster's redshift. This allows the model to have only a few free parameters for all of the cluster members since positions (and sometimes ellipticity/position angle) are informed by the light distribution. Large-scale halos, on the other hand, are usually allowed to vary freely. Both kinds of halos are parametrized by given density profiles. Free-form models, on the other hand, do not put such constraints on the mass of the halos. This freedom is both useful in that it can capture oddities in the mass distribution, but can also be a disadvantage if there are less constraints than free parameters. Hybrid techniques are those which have free-form large-scale halos, but use given density profiles for small-scale halos. 

\subsection{Modeling Deliverables}
Each team submitted a number of deliverables for their fiducial model, as well as a number of realizations of the model. These realizations, which we will refer to in this work as `range maps', varied from 40 to over 200 and were meant to sample the uncertainty in a model. It is important to note that lensing quantities depend on the distances between the observer, lens, and source. For the range maps, all teams submitted shear ($\hat{\gamma}$) and convergence ($\hat{\kappa}$), or surface mass density, maps that correspond to a source at infinite distance. From there, the quantities can be found at any source redshift using 
\begin{equation}
\kappa = \frac{D_{ls}}{D_s}\hat{\kappa}\;\; ; \;\gamma = \frac{D_{ls}}{D_s}\hat{\gamma},
\end{equation}
where $D_{s}$ and $D_{ls}$ represent angular-diameter distances from the observer to the source and from the lens to the source, respectively. 
 
Further, while the fiducial model submitted had to include magnification maps for $z=1,2,4$ and $9$, one can find the magnification at any redshift by using
 \begin{equation}
\mu =  \frac{1}{(1-\kappa)^2-\gamma^2}.
\end{equation}
 We note that these equations are only true for a 2D model, i.e. a single lens plane. With a 3D model, there are multiple lens planes and thus the shear and convergence are not so easily scaled \citep[see e.g.][]{schneider1992}. 

One can also use these $\hat{\kappa}$ maps to find the mass predicted by a model. The convergence is defined as the surface mass density divided by a critical surface density:
\begin{equation}
\hat{\kappa} = \frac{\Sigma}{\hat{\Sigma}_{crit}}, \mathrm{where}\;\; \hat{\Sigma}_{crit} = \frac{c^2D_l}{4\pi G},
\end{equation}
and $D_l$ is the angular-diameter distance from the observer to the lens. By summing the convergence, for example in circular apertures as we do in this work, the mass can be computed.

\subsection{Participating Teams}
In this work, we consider models from five teams using parametric methods, two using free-form methods, and one using a hybrid technique. Three teams (Caminha, CATS, and Sharon) share the same modeling code (Lenstool), while all other teams use separate codes. For an in-depth overview of the techniques for each team, we point the reader to \cite{meneghetti2017} or \cite{priewe2017}.

The teams using parametric methods are Caminha \citep{caminha2017}, Clusters As TelescopeS (CATS) \citep{jullo2007,jauzac2012,jauzac2014,richard2014}, Glafic \citep{oguri2010,ishigaki2015,kawamata2016,kawamata2018}, Keeton \citep{raney2019}, and Sharon \citep{jullo2007,johnson2014}. 

Two teams use free-form methods: Brada\v{c}/Strait \citep[shortened to Brada\v{c} in plots for space;][]{bradac2005,bradac2009,strait2018} and Williams \citep{liesenborgs2007,mohammed2014,grillo2015}. One team, Diego, use a free-form method but assumes that light traces mass for the galaxies, i.e. each galaxy is initally assigned mass based on its surface brightness and later optimized \citep{diego2005a,diego2005b,diego2007,diego2015,vega2019}. 

All teams use only strong lensing constraints except Brad\v{c}/Strait, who also employ weak lensing. The number of halos (large-scale and galactic) varies among the teams, as do the density profiles of the halos for the parametric models. The number of images used as constraints can differ as well and, in some fields, by large amounts (e.g. $\sim$100 images). 

\section{Mass Comparison}
\label{sec:mass-comp}
\subsection{Overview}
One of the ways that we can compare the results from all teams is by looking at mass profiles. The goal of a lens model is to find the underlying mass distribution and source configuration that can produce the lensed images seen in the data. This is not an easy task, especially in cluster lensing due to the inherent complexity of galaxy clusters. Further, the clusters that are most likely to be chosen as cosmic telescopes are those that are both large on the sky and very massive. These two factors combine to give a larger area on the sky where background galaxies can be strongly lensed. However, this can cause a selection bias for clusters that are undergoing a merger, which can increase both the density and physical size of a cluster. A configuration that is also favorable to lensing many images is multiple large-scale halos along the line of sight, which can boost lensing strength \citep{wong2012,bayliss2014}.

It can be difficult for lens models to differentiate between mass profiles in a cluster using just image positions as constraints. For example, a recent study of the Hubble Frontier Field MACS J0717.5+3745 found that the data fit models with cored and non-cored dark matter halos equally well, even with 132 constraints \citep{limousin2016}. This is also seen in mock data: a model with isothermal halos can fit position data just as well as a model with NFW halos even though the density profiles are obviously different, as are the resulting image magnifications \citep{shu2008}. 

A common metric used to compare mass distributions found by lens modeling is the 1D mass profile. This was used in \citet{meneghetti2017} to compare the results from multiple teams modeling two mock clusters, as a way of determining how accurate and precise the models were. It was found that, though the multiple teams used different density profiles for the halos and different modeling techniques, they were able to recover 1D mass profiles to within 15\% of the true value. 

In this work, we do not know the true mass distribution of the cluster, but it is still useful to compare the mass profiles obtained by the different modeling teams and see the extent to which they agree or disagree. We construct our 1D profiles by computing the mass in circular apertures centered on the BCG. In the following subsections, we give a brief introduction to each lensing field, including a sky map. This map includes two solid circles at 5 and 100 arcsec from the BCG, which correspond to the $x$-axis limits of the 1D mass profiles, shown in the right panels. The profiles are split between parametric (top) and free-form (bottom) techniques for clarity. We note that, for each model, we plot the 1D profiles for all of the submitted range maps, such that the thickness of the line illustrates the uncertainty in the model. We also note that the lines very often overlap. The median across all models and realizations is plotted in black on both panels for reference. 

Since the modeling teams were allowed to choose the size of their maps, the mass profiles do not all go out to the same radii. Further, parametric models use certain density profiles for their halos, thus mass continues to grow at large radii. Free-form techniques, on the other hand, have different priors and regularizations. This can produce flatter profiles at larger radii where there are no lensed image constraints and, as we will see in Sec. 4, lower magnifications. We also note that, though we have created both 2D and 3D models of each field, we only include the 2D models in the mass analysis. A multi-plane model has mass at different redshifts and thus is not a fair comparison to single-plane models. 

We indicate the locations of lensing constraints in two ways. The dashed circle in the sky map indicates the spatial extent of the lensed images we used in our models \citep[see][]{raney2019}, which are primarily the \textsc{gold} sample. In many cases, the images are not centered around the BCG because its position does not coincide with the center of the mass distribution due to merging systems. We also mark the image positions as vertical lines in the top mass profile panel. This helps to show the distribution of these images and informs where the models might be most tightly constrained. We stress that the sample shown is unique to our team and fairly conservative since it is primarily restricted to images with spectroscopic information. Other teams may have used different images, and thus their models will be constrained differently.

\subsection{Abell 2744}

\begin{figure*}
\centering
\includegraphics[width=\textwidth]{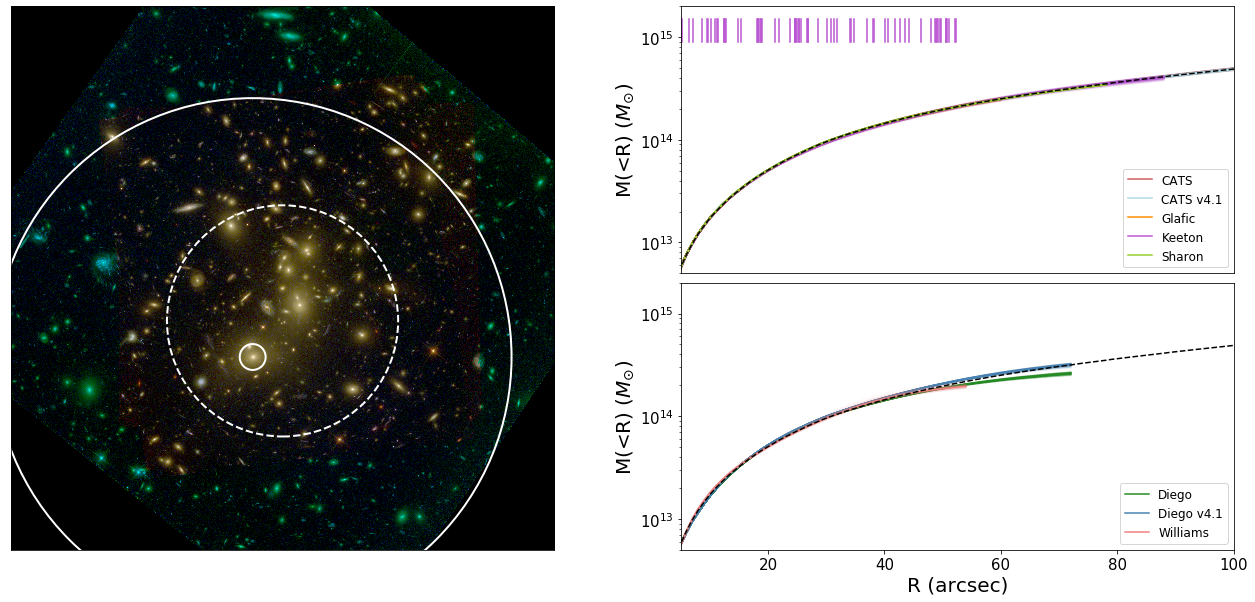}
\caption{\textit{Left:} HST multi-band color image \citep[produced using Trilogy,][]{coe2012} of the cluster Abell 2744. Solid circles are centered on the BCG and have 5 and 100 arcsec radii, corresponding to the $x$-axis limits in the panels on the right. The dashed circle encloses the majority of spectroscopically confirmed images: specifically, those images we used in our own modeling \citep[see Appendix B in][]{raney2019}. The panel is 3.5 arcmin on a side and is oriented such that North is up and East is left. We note that the size of the panel does not correspond to the size of the submitted maps of the teams. \textit{Right:} Mass profiles centered on the BCG and circularly averaged, computed from the publicly available $\hat{\kappa}$ maps for a source at infinity. All submitted maps are plotted, including the realizations such that the thickness of the line describes the error. The median profile across all teams is also plotted (black, dashed). Models employing parametric techniques are shown on top while free-form/hybrid models are in the bottom panel. We note that some submitted maps covered a smaller area than others, causing the profiles to truncate at different radii. We also include lines indicating the distance of images from the BCG for the constraints used in our model.}
\label{fig:a2744-skymap}
\end{figure*}

This field, part of the Abell galaxy cluster catalog \citep{abell1989}, was the first HFF cluster to be observed by the \textit{Hubble Space Telescope} (\textit{HST}) and has a redshift of $z=0.308$. It is a system undergoing a merger, as evidenced by a number of factors. The first is that the cluster is physically very large. In Fig. \ref{fig:a2744-skymap} we show the HST color image of the field, but we note that the cluster extends to the northwest, past the field of view (FOV). In the figure, we see two galaxies with similar brightness, which could both be classed as BCGs; $\sim$2 arcmin away, there are three more galaxies with the same brightness down to photometric errors \citep{mann2012}. However, since there are more cluster member galaxies around the southern two BCGs, this is considered the main part of the cluster. 

Optical and X-ray studies suggest that the system has undergone two mergers in the recent past, one of which was line-of-sight \citep{kempner2004,owers2011,merten2011}. This would explain both the large number of BCGs and the offsets found between peaks in the X-ray data and the positions of the cluster members. While the mass outside the HST FOV is affecting the lensing on some scale, it is not well constrained due to the lack of lensed images in that region, far from the southern core. Most modeling teams did find that the models preferred to place a large-scale halo to the northwest of the main cluster, as we will see in the magnification maps in the next section. The image constraints in the main part of the cluster are fairly numerous: around 70 images have spectroscopic redshifts. This is in large part due to a recent spectroscopic survey \citep{mahler2018} using MUSE. 

In the latest round of modeling, six teams created models of the field. We show the 1D mass profiles for each model in the right hand panels of Fig. \ref{fig:a2744-skymap}. In the top panel, we show models that were made using a parametric method, while those shown in the bottom panel were made with free-form methods. The width of the line represents the scatter in the model using the submitted range maps. Some teams cut off before the edge of the plot due to smaller area of their submitted maps. 

It is immediately obvious that all the models agree fairly well. The two Diego models, which here differ in their constraint selection (\textsc{gold+silver+bronze} vs. \textsc{gold}), are fairly different at larger radii: the v4.1 profile agrees with the parametric and median curves, while the v4 profile has a shallower slope. We will see in Sec. 4 that the magnifications maps of these two models are also quite different. Nonetheless, the scatter among all models is surprisingly low with 1$\sigma$ scatter of $<5\%$ out to an arcminute from the BCG. In fact, the scatter becomes $<1\%$ at 14 arcsec from the BCG, the lowest value out of all the fields in the HFF sample.

\subsection{MACS J0416.1-2403}

\begin{figure*}
\centering
\includegraphics[width=\textwidth]{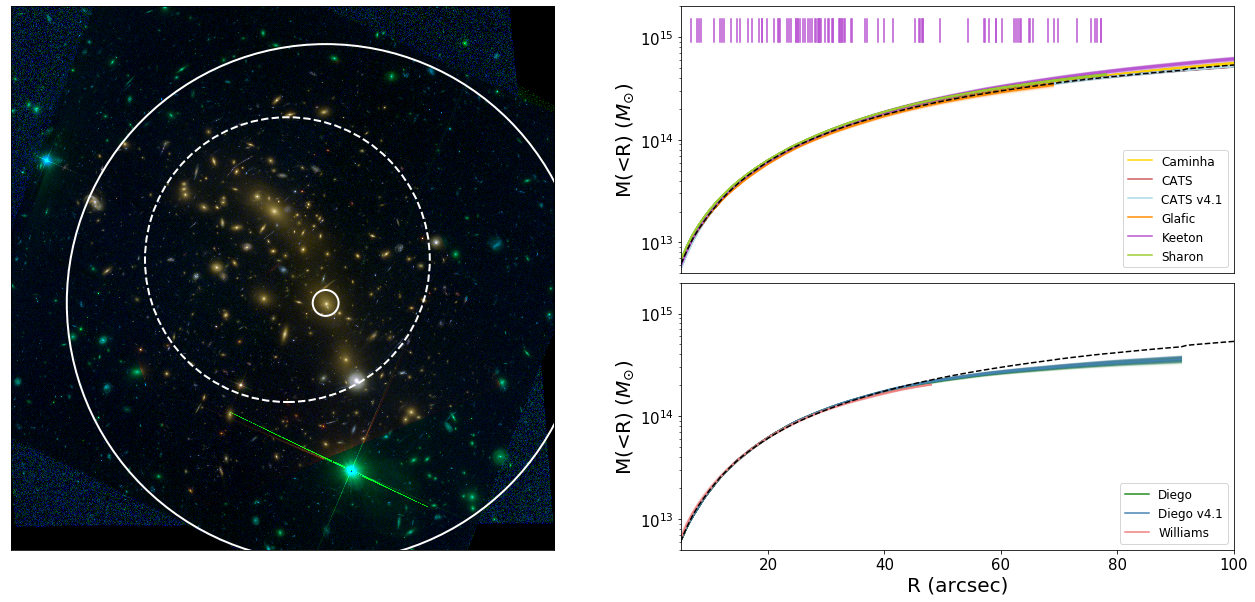}
\caption{Similar to Fig. \ref{fig:a2744-skymap} for MACS J0416.}
\label{fig:m0416-skymap}
\end{figure*}

The second of the Hubble Frontier Fields to be observed by \textit{HST} is this cluster at $z=0.396$ from the Massive Cluster Survey \citep[MACS;][]{ebeling2001}. Similar to Abell 2744, there is evidence that it is undergoing a merger, though one that is not quite as dramatic. From the sky map in Fig. \ref{fig:m0416-skymap}, one can see that there are two BCGs with similar brightness. Further, the X-ray map is distinctly doubly-peaked \citep{mann2012}. The merger is likely one that is along the line-of-sight. Due to this orientation, the lensing area is elongated in such a way to produce a large number of triple images in a ladder configuration.  

Indeed, this field has the most images in the \textsc{gold} sample out of all the six fields: $\sim\!95$. This allows for models that can be well constrained, which is indeed what we see in the right hand panels of Fig. \ref{fig:m0416-skymap}. The mass profiles are very similar, specifically at radii between 10 and 40 arcsec, where the scatter is around 2.5\%. It is not surprising that this is also the range in which the bulk of the images are found. It is interesting to note that there is more scatter at larger radii in this field than in Abell 2744, where the mass distribution is known to extend beyond the modeled area.

\subsection{MACS J0717.5+3745}
\begin{figure*}
\centering
\includegraphics[width=\textwidth]{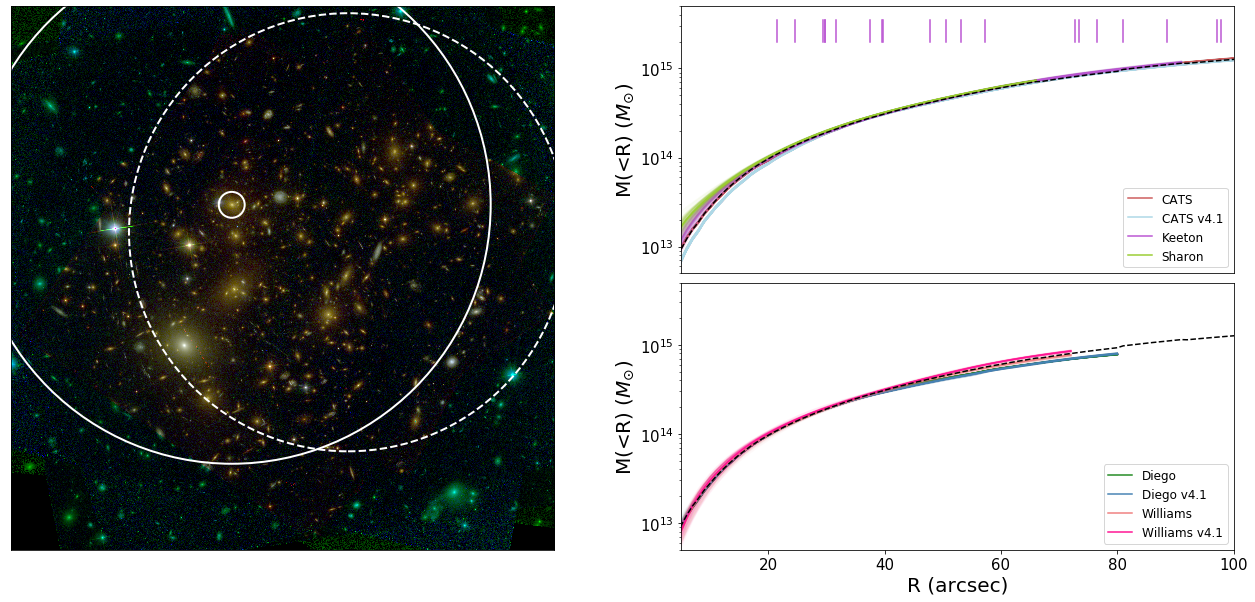}
\caption{Similar to Fig. \ref{fig:a2744-skymap} for MACS J0717.}
\label{fig:m0717-skymap}
\end{figure*}

This cluster, found as part of the MACS survey \citep{ebeling2001}, is superlative among the HFF sample in many respects. It has the highest redshift at $z=0.545$, slightly higher than MACS J1149 at $z=0.543$. It is the most massive cluster in the sample and also likely the most complicated; it was considered the most disturbed system at $z>0.5$ due to the complex nature of its X-ray data \citep{ebeling2007}. Part of the complexity comes from a filament \citep{ebeling2004,jauzac2012}, which could be causing the odd elongated nature of the lensing critical curves that we will see in the next section.

We see in Fig. \ref{fig:m0717-skymap} that the field is not a typical cluster with a BCG in the center of smaller cluster member galaxies. Indeed, the galaxy classed as the BCG (within the smallest circle in the figure) is at the center of neither the cluster members nor the area covered by lensed images (shown by the dashed circle). The proposed filament can be seen in the figure as the swath of cluster galaxies extending to the upper right. We note that the bright galaxy to the bottom left is likely a foreground galaxy based on a photometric redshift of $z=0.155\pm0.03$ from CLASH \citep{postman2012,molino2017} and Subaru/Suprimecame imaging \citep{medezinski2013}. Yet another source of complication comes in the form of a possible LOS structure in the field for which \citet{williams2017} found evidence.

Unfortunately, this complex cluster also has the least number of spectroscopically confirmed images with which its mass can be constrained: less than 30. That is not to say the field lacks candidate images; the CATS team, for example, used 132 images in their v4 and v4.1 models. These two models are different in that they have either cored or non-cored halos, respectively. Even with the large number of constraints they used, they found that both models were able to fit the data equally well \citep{limousin2016}.

This is evident in the mass profiles shown in the right hand panels of Fig. \ref{fig:m0717-skymap}, where the CATS v4 and v4.1 models (red, blue) do indeed disagree at low radii. Interestingly, there does not appear to be a lot of intrinsic scatter in each model. This is not true for the other two parametric teams; the Sharon team's model has a fairly large spread around the core of the cluster, as does our model. All of these models converge at higher radii, though, which is unsurprising: the constraints also extend to a large radius.  

For the free-form teams, the results are slightly different. There is some scatter at smaller radii, but not as much as among the parametric models. Further, there is more scatter at larger radii. The two Williams models are different, but do straddle the median curve. The two Diego models, on the other hand, agree with each other very well, but lie the farthest from the median profile.

\subsection{MACS J1149.4+2223}
\begin{figure*}
\centering
\includegraphics[width=\textwidth]{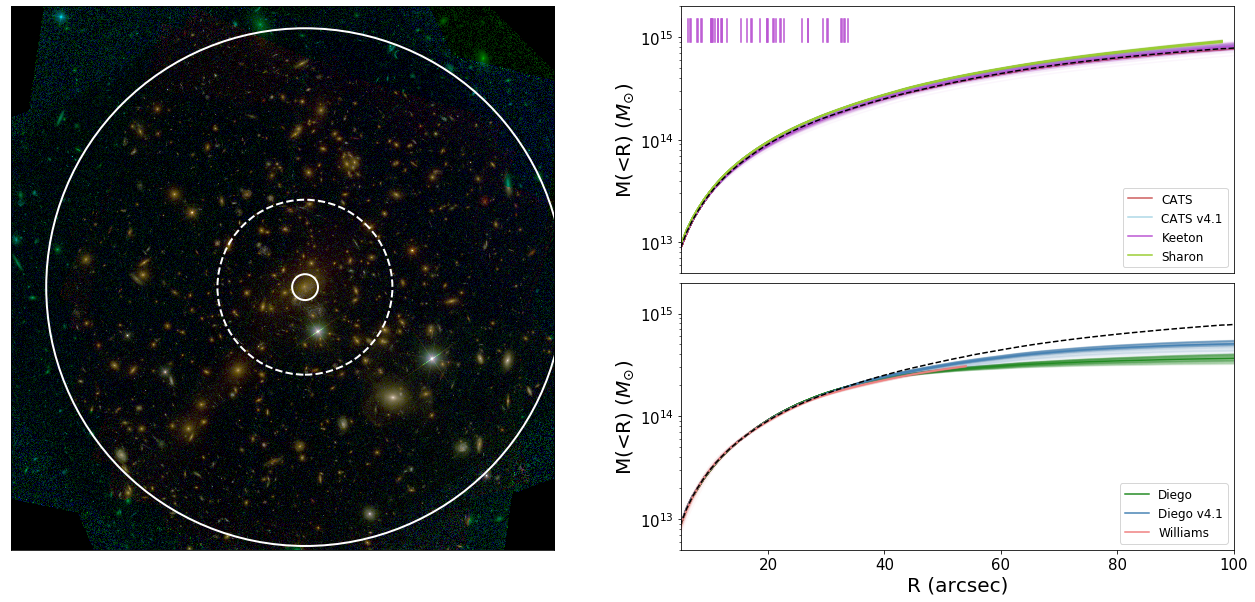}
\caption{Similar to Fig. \ref{fig:a2744-skymap} for MACS J1149.}
\label{fig:m1149-skymap}
\end{figure*}

This cluster is also at a fairly high redshift ($z=0.543$), but is less complex than MACS J0717. For example, the BCG is notably brighter than any other galaxy in the field and lies nicely at the epicenter of the lensed images, as seen in Fig. \ref{fig:m1149-skymap}. It does have a somewhat elongated mass distribution so it is likely undergoing a merger, but one that is in later stages than some of the other fields. 

The cluster has been the focus of many studies, due in large part to a triply-imaged spiral galaxy. Two of its images sit close to the BCG, the closer of which is fairly distorted. The second image has a spiral arm further lensed into an Einstein-cross configuration by a cluster member galaxy, but otherwise shows only a small amount of distortion. The third image, $\sim\!$20 arcsec from the BCG, is also mostly intact. These three images can thus be used to give constraints on the mass distribution of the BCG and cluster core \citep[e.g.][]{zitrin2009c,rau2014}. 

This spiral galaxy was also the host of SN Refsdal, a Type II supernova that was found in the arm of the galaxy that was further lensed by a cluster member \citep{rodney2016,treu2016}. The four images of the SN in the cross configuration were named S1-S4. The SN was also set to appear in the image of the galaxy closest to the cluster core, but not for a time after S1-S4. Thus this other image (SX) could be used as a test of the predictive abilities of lens models. The models were able to predict the position of SX quite well but its time delay, i.e. when it would appear, proved harder to pin down \citep{kelly2016}. Still, the ability to make somewhat accurate predictions is a good sign that the lens modeling is headed in the right direction.

It is important to note that, while this field was the subject of many studies, there are still relatively few lensed images with spectroscopic redshifts; only 22 images from 9 sources were ranked \textsc{gold}. Star forming knots within the spiral arms of the Refsdal host galaxy \citep[e.g. see][]{kawamata2016} can be used as further constraints on the model. We do note, however, that two of the images of this galaxy are $<10$ arcsec from the BCG, and thus the majority of the constraints are on the inner region of the cluster. 

Given that there are many constraints close to the BCG and so few farther out, it is then unsurprising that the mass profiles shown in the right hand panels of Fig. \ref{fig:m1149-skymap} are tight at small radii but become broader as radius increases. In fact, at an arcminute from the BCG, MACS J1149 has the highest scatter across the six fields: 20\%. This is driven by the differences between the two Diego models, which are not only very different from each other, but are also quite far from the median and the parametric models.

\subsection{Abell S1063}
\begin{figure*}
\centering
\includegraphics[width=\textwidth]{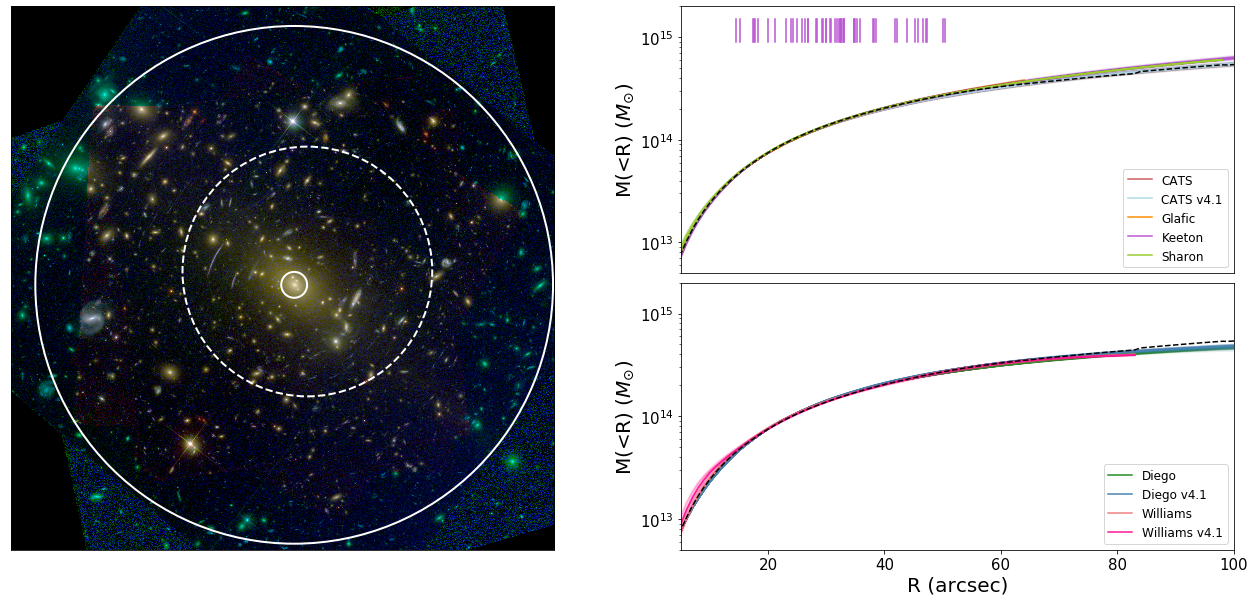}
\caption{Similar to Fig. \ref{fig:a2744-skymap} for Abell S1063.}
\label{fig:a1063-skymap}
\end{figure*}

This cluster ($z=0.348$) is the most well-behaved in the HFF sample. For example, there is one clear BCG which lies at the center of the cluster galaxies, as shown in Fig \ref{fig:a1063-skymap}. It can also be seen that the \textsc{gold} sample, consisting of almost 50 images from 19 sources, are mostly clustered around the BCG, though there are quite a few candidate images to the northeast. There is evidence that the system is undergoing a merger based on dynamical studies \citep{gomez2012}, which could explain this. Nonetheless, it is not as dramatic of a merger, or perhaps is in a later stage than other clusters in the sample. 

The mass distribution of the cluster core is well-constrained among the parametric models, though the free-form/hybrid models show scatter some at low radii. Like the Diego models, the two Williams models differ in their constraints: in this case, v4.1 is only the \textsc{gold} sample, while v4 uses \textsc{gold+silver+bronze}. It is interesting that the largest differences are seen near the core of the cluster in the free-form models while parametric models show more (though still a small amount of) scatter at larger radii. Nonetheless, the mass of the cluster is very well constrained with 1$\sigma$ scatter of less than 5\% past 10 arcsec. 

\subsection{Abell 370}
\begin{figure*}
\centering
\includegraphics[width=\textwidth]{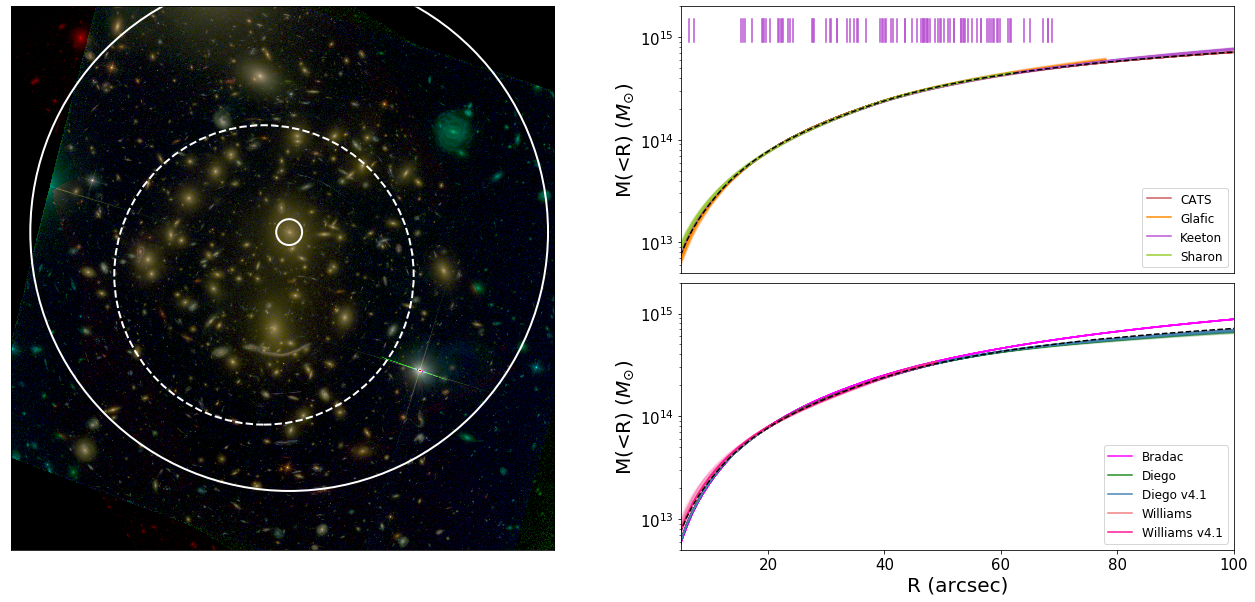}
\caption{Similar to Fig. \ref{fig:a2744-skymap} for Abell 370.}
\label{fig:a370-skymap}
\end{figure*}

This cluster ($z=0.375$) was the first in which a strongly lensed galaxy was discovered, stretched into a giant arc \citep{soucail1987,lynds1989}. In the 30 years since it was found, it has been the subject of many studies on both weak and strong lensing \citep[e.g][etc.]{Abdelsalam1998,Bezecourt1999,Medezinski2010}. Its structure and galaxies have also been studied \citep{defilippis2005,lah2009}, pointing towards a system undergoing a line-of-sight merger. In Fig. \ref{fig:a370-skymap}, we see evidence of this complexity: two possible BCGs and a large area over which lensed images are found. Indeed, this field has the second highest number of spectroscopically-confirmed images, in part due to a recent MUSE survey \citep{lagattuta2017,lagattuta2019}.

The large number of constraints on the field from the 90 \textsc{gold} images does seem to be able to combat the complexity, at least for the parametric models. The mass profiles of Fig. \ref{fig:a370-skymap} show scatter at small radii, but most of the models agree very well at larger radii. Indeed, this field has the smallest 1$\sigma$ scatter at an arcminute from the BCG out of all six fields: only 2\%. This is probably due in part to the wide area over which the image constraints are spread, similar to what was seen in MACS J0717 but with many more images. 

However, it is interesting to look at the outlier case of the Brada\v{c}-Strait model, which is significantly higher than the other mass profiles. Recall, this team also employed weak lensing data in addition to the strongly lensed images to constrain the mass distribution at larger radii; none of the other teams did this. It is unclear whether this higher mass profile stems from the weak lensing alone, or also from their modeling methodology, but it is an interesting result.

\section{Magnification Comparison}

\subsection{Overview}
In order to determine the intrinsic properties of a lensed galaxy, the amount of magnification must first be determined. This makes magnification the most important quantity in the search for and study of high redshift galaxies, but it can also be hard to constrain. It is highly nonlinear and a small change in model parameters can produce large changes in the magnification at a specific point, especially if it is close to the critical curves (defined as where $\mu\rightarrow\infty$).

In this section, we seek to compare the magnification maps submitted by each team. To do this, we first find the largest area in common between the range maps of all teams and trim the maps to this area; this does sometimes exclude interesting regions that the team(s) with the smallest area did not model, but it is necessary to make a fair comparison overall. We then find the lowest spatial resolution, i.e. highest area per pixel, among the teams and resample all of the maps to this resolution. We use 2-d linear interpolation to find values at the same locations in each map instead of rounding to the nearest pixel in order to prevent artifacts, specifically in the 2-d histograms. 

This yields a data cube comprised of the range maps, now with a common area and resolution, for each model. It is not straightforward to analyze such a dataset; we would want something that incorporates the errors, but also does not ignore the spatial aspect of the maps. \citet{priewe2017} tackled this problem in various ways for version 3 models of Abell 2744 and MACS J0416, namely looking at $\sim\!200$ pixels set in a grid around the cluster core and analyzing the spread in magnification histograms for various magnification bins. While this accomplished the goal of showing the increasing spread of magnifications across the field, the spatial context was mostly lost. That is, if one part of the map showed a higher spread than other parts (say, due to an interloping foreground galaxy), this would not be apparent in a magnification histogram. 

\subsubsection{Half-sample mode}
We analyze the models in two separate ways, the first of which uses half-sample mode (HSM) maps. The half-sample mode is a robust way to find the value of maximum likelihood of a random variable. This method finds the peak of a histogram which may be non-Gaussian and/or have outliers. It is important that the estimator used be robust to outliers; near the critical curves, small shifts in the mass distribution can cause large shifts in magnifications. 

The estimator is found by a recursive method which cuts a sample down to the smallest interval that encloses half of the data until the mode is found \citep[see][]{bickel2005}. This is done for each model, pixel by pixel, across the range of all realizations for that model. In this way, a data cube is condensed to a single map, but errors are still somewhat included. We can then show both the HSM map and a ratio between two HSM maps of different teams to highlight variations. 

We note that the HSM technique does introduce a ``fuzziness'' artifact in the maps, specifically with models that show significant scatter. For example, our models include scatter in the mass-luminosity relation. This causes variations in the critical curves around the galaxies, which can manifest as washed out features in the HSM maps. It is also a prevalent feature in the free-form maps of the Williams team. However, this is also useful: their maps are particularly free outside of the strong lensing region due to the freedom in their methodology, but fairly well constrained within this region, which is highlighted by the HSM maps. 

\subsubsection{2-d histograms}
Another way to visualize the difference between the models is a 2-dimensional histogram. With it, we depict the joint probability distribution $P(\mu_1,\mu_2)$ that model 1 predicts $\mu_1$ and model 2 predicts $\mu_2$ taken across all pixels and between 1000 pairs of maps sampling the range.  This is particularly useful in that we naturally sample from the complete set of realizations and thus get a sense of the full range of the models. Since the maps are $\sim\!250$ pixels on a side, the histograms then have roughly $7\times10^7$ pixels in the 1000 pairs. We note that many teams have 100 or more realizations of each model, thus the 1000 pairs undersample the full suite, but the results show little to no change if the number of pairs is increased. 

It is easy to pick out by eye which models are relatively similar to each other in a 2-d histogram. Models with many pixels in common will show high density along the one-to-one line with varying scatter depending on how tightly constrained the parameters are in a given model; if they are not tightly constrained, they fall in a cloud around $\mu_y=\mu_x$. Other differences in the models can result in more interesting features in the 2-d histograms. For example, if a model has bimodal characteristics and the realizations fall within two classes, this might appear in the 2-d histogram as another track of relatively high density, as opposed to a cloud due to scatter.

\subsubsection{Presentation of results}
In the following subsections, we first show the HSM magnification maps for the area in common for all of the models. This allows us to look at broad stroke similarities and differences, and to compare the overall shape of the models. We try to keep a common structure to the plots for the fields, but there will be some variations due to some teams not modeling all of the fields. The second plot for each cluster shows both the HSM ratio maps and the 2-d histograms for easy comparison. The ratio panels are arrayed such that the HSM of the team denoted on the $x$-axis is divided by that of the team on the $y$-axis. Thus a panel showing mostly red, i.e. positive ratios, indicates that the magnifications in the model of the team on the $x$-axis are higher than those of the team on the y-axis. 

The 2-d histograms fill in the rest of the space left from the ratio map triangle plot nicely. It offers the same combinations of model comparisons, except transposed: e.g. the left-most column corresponds to the bottom row. Having these plots next to each other is quite useful: areas of red, positive values in the spatial maps correspond to the area above the one-to-one line in the 2-d histogram. The 2-d histograms along the diagonal from the bottom left to the top right show self-comparisons, i.e. both datasets making up the 1000 pairs of realizations come from the same model. This allows us to see what the statistical scatter of a given model is and compare it to the scatter seen among the teams.

\subsection{Abell 2744}

\begin{figure*}
\centering
\includegraphics[width=0.75\textwidth]{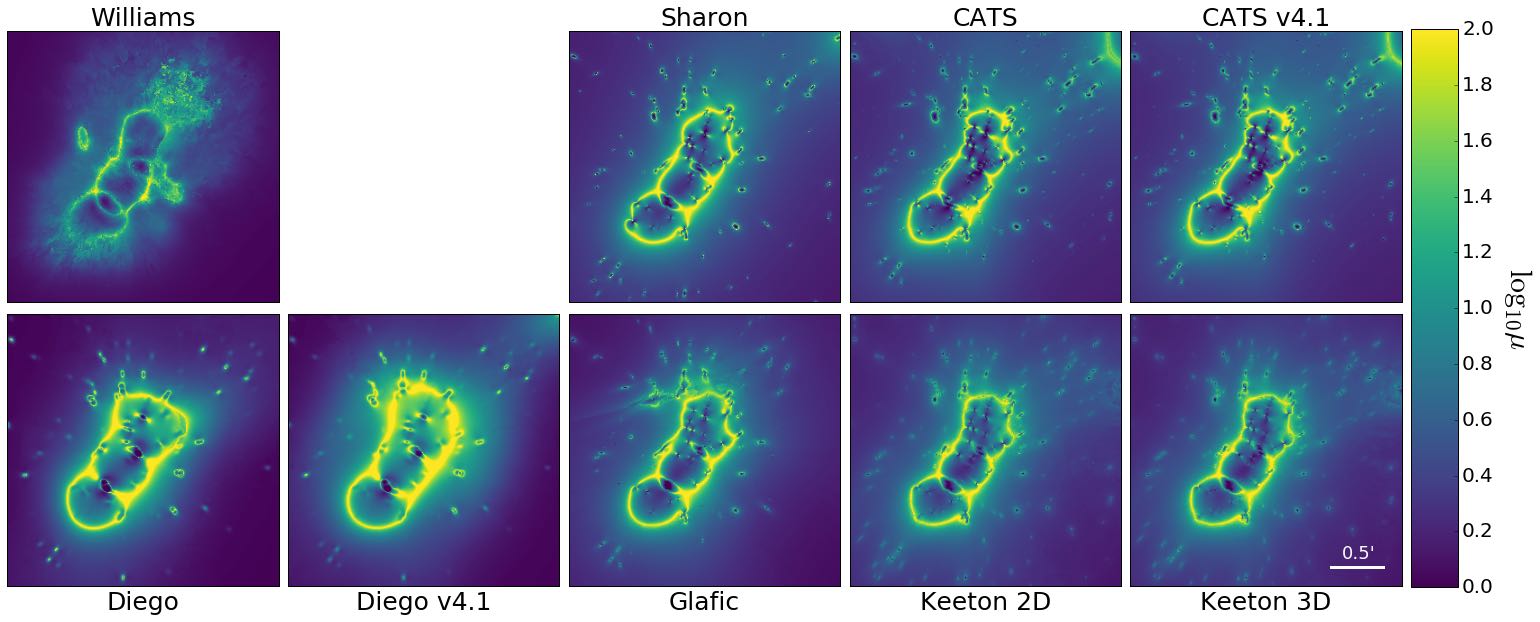}
\caption{The half-sample mode (HSM) magnification maps for a source at redshift $z=9$ from the suite of realizations for each model of Abell 2744; each panel is version 4 unless specified otherwise. Every plot covers the same area and is oriented such that up is North and left is East. The overall shape of the critical curves is seen to be mostly consistent between models, but a number of differences exist.}
\label{fig:a2744-hsm}
\end{figure*}

\begin{figure*}
\centering
\includegraphics[width=0.9\textwidth]{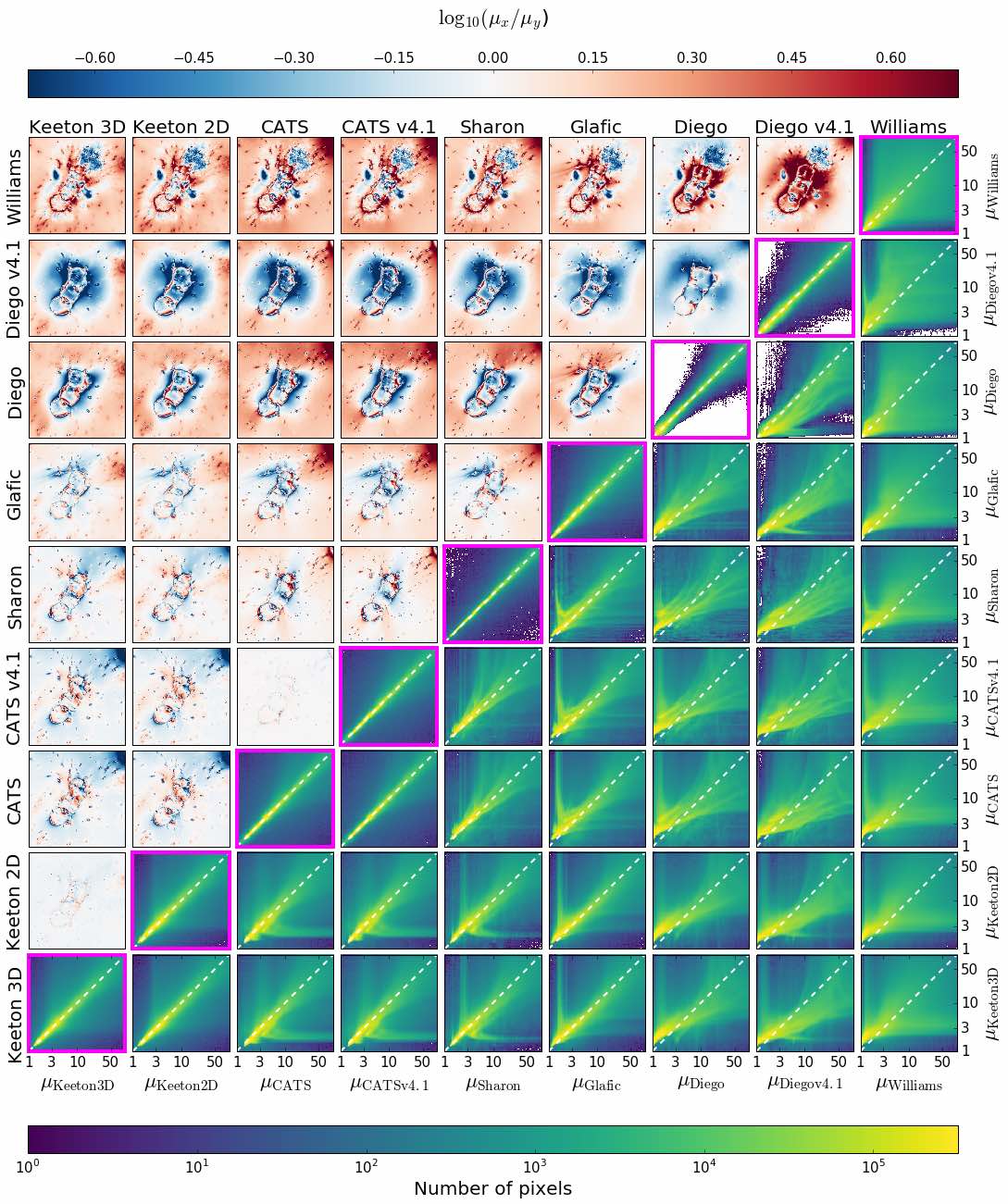}
\caption{\emph{Upper left:} Ratios of HSM maps for every pair of models of Abell 2744, arranged as Model$_x$/Model$_y$ such that, e.g., the first row is all models divided by the Williams model. Note the color scale: we use logarithmic values due to the wide range in magnifications. \emph{Lower right:} Two-dimensional histograms showing the probability distribution $P(\mu_1,\mu_2)$ that Model$_x$ predicts $\mu_1$ and Model$_y$ predicted $\mu_2$. Note that, for each panel, this is calculated across 1000 pairs of models drawn from the submitted realization maps. Models that are very similar to one another will have a high density along the one-to-one line (white, dashed). Model self-comparisons, i.e. a model vs. itself, are plotted along the diagonal and outlined in magenta. The various structures seen in the plots can be explained by differences in mass structures in the models, as described in the text. All panels assume a source redshift of $z=9$.}
\label{fig:a2744-comp}
\end{figure*}

Six teams produced nine models of this field. The HSM maps of each model are shown in Fig. \ref{fig:a2744-hsm}. For this field, the difference between the two CATS models is the same as that between the Diego models: v4 used only \textsc{gold} constraints, while v4.1 used \textsc{gold+silver+bronze}. 

Based on the HSM maps shown in Fig. \ref{fig:a2744-hsm}, all models seem relatively consistent, especially near the core of cluster. This is where one would expect them to be most similar since it is where the bulk of the images are. Some form of a double band structure can be seen in all of the models, caused by the two bright, large cluster members seen in Fig. \ref{fig:a2744-skymap}, whose influence is important enough to be captured by the free-form models. 

These similarities are encouraging, but there are also clear differences. For example, some models have a halo off to the northwest (upper right in Fig. \ref{fig:a2744-hsm}), which the Williams, Diego, and Glafic models do not require. This halo does not seem to be in much agreement among the models which do have it. The two CATS models have a halo with a large critical curve while the models of Sharon and Diego v4.1 prefer a halo with smaller critical curve. Our two models both place a halo in this region with similar Einstein radius, though the 3D model finds one that is more diffuse, sometimes not even producing a critical curve. Both of our models place the halo due west of the top of the cluster critical curve while the other teams put it to the northwest. We find that halo to have a wider range in parameters than the other two large-scale halos, which causes the blurry edges seen in the HSM map. The two CATS models also put in another halo to the northeast (upper left) which is cut off in the maps shown here. The Glafic model appears to sometimes have a quite elongated halo near one of the galaxies to the east of the top of the cluster critical curve. 

To see how these differences compare quantitatively, we show ratio maps in Fig. \ref{fig:a2744-comp}. Immediately, a number of trends can be seen. The free-form vs. parametric model comparisons at the upper left of the figure all seem to have a red base, even away from the cluster core. This is due to the free-form models having lower magnifications away from, even if there are higher magnifications near, the core of the cluster, as shown by the Diego models. Recall that in the mass profiles, the Diego v4 and Williams models were lower than the median profile. While the Diego v4.1 model agreed very well with the median profile and those of the parametric models, it is clear this added mass causes the magnification maps to look very different. 

The parametric models show slightly less variations, though the halos outside of the core affects the ratios. It can be seen in the Sharon vs. CATS and CATS v4.1 panels that, though both teams predict a halo to the northwest, there is disagreement in its parameters. It is clear that the Glafic model has no halo to the west of the cluster, and their elongated halo to the east stands out more clearly here than in Fig. \ref{fig:a2744-hsm}. 

To determine how the full suites of realizations compare among the models, we also show the 2-d histograms in Fig. \ref{fig:a2744-comp}. We see that two models that were very similar in the ratio maps, e.g. CATS v4 and v4.1, produce a 2-d histogram that is heavily populated, as expected, along the one-to-one line (dashed white). Some comparisons do not fall along the one-to-one line at all, e.g. Diego 4.1 vs. CATS v4.1; others may vaguely fall along this line, but have large spreads, e.g. Williams vs. Diego. 

The structures that appear in these panels are also informative about the models themselves. For example, in the Sharon vs. Keeton 2D 2-d histogram panel, there are horizontal and vertical branches that correspond to the extra halos that the two models include. If only one model has a halo at a certain position, then the model without the halo will have constant low magnification, while the other model will show increasing magnifications around the critical curves. Since the Keeton and Sharon teams have both halos, but in different places, this creates two branches.

These plots are also important in that they show that, even at low magnifications, the models do not necessarily agree. The parametric vs. parametric panels are well-populated around $\mu_y=\mu_x$ at low magnifications, but this is not true for the free-form vs. parametric models. This is not surprising given what we see in the ratio panels; it is also important to point out that much of this is caused by the region outside of the critical curves. 

\subsection{MACS J0416.1-2403} \label{ssec:m0416}

\begin{figure*}
\centering
\includegraphics[width=0.7\textwidth]{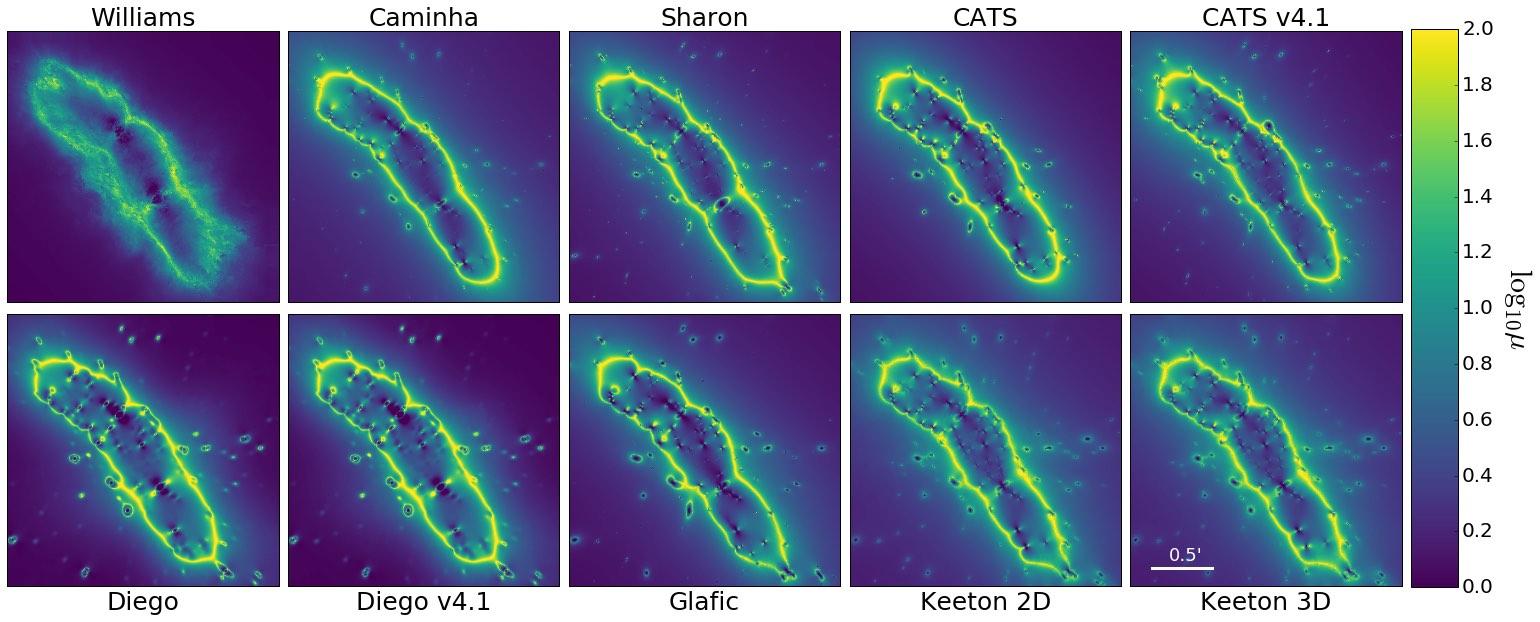}
\caption{Similar to Fig. \ref{fig:a2744-hsm} for MACS J0416.}
\label{fig:m0416-hsm}
\end{figure*}

\begin{figure*}
\centering
\includegraphics[width=0.9\textwidth]{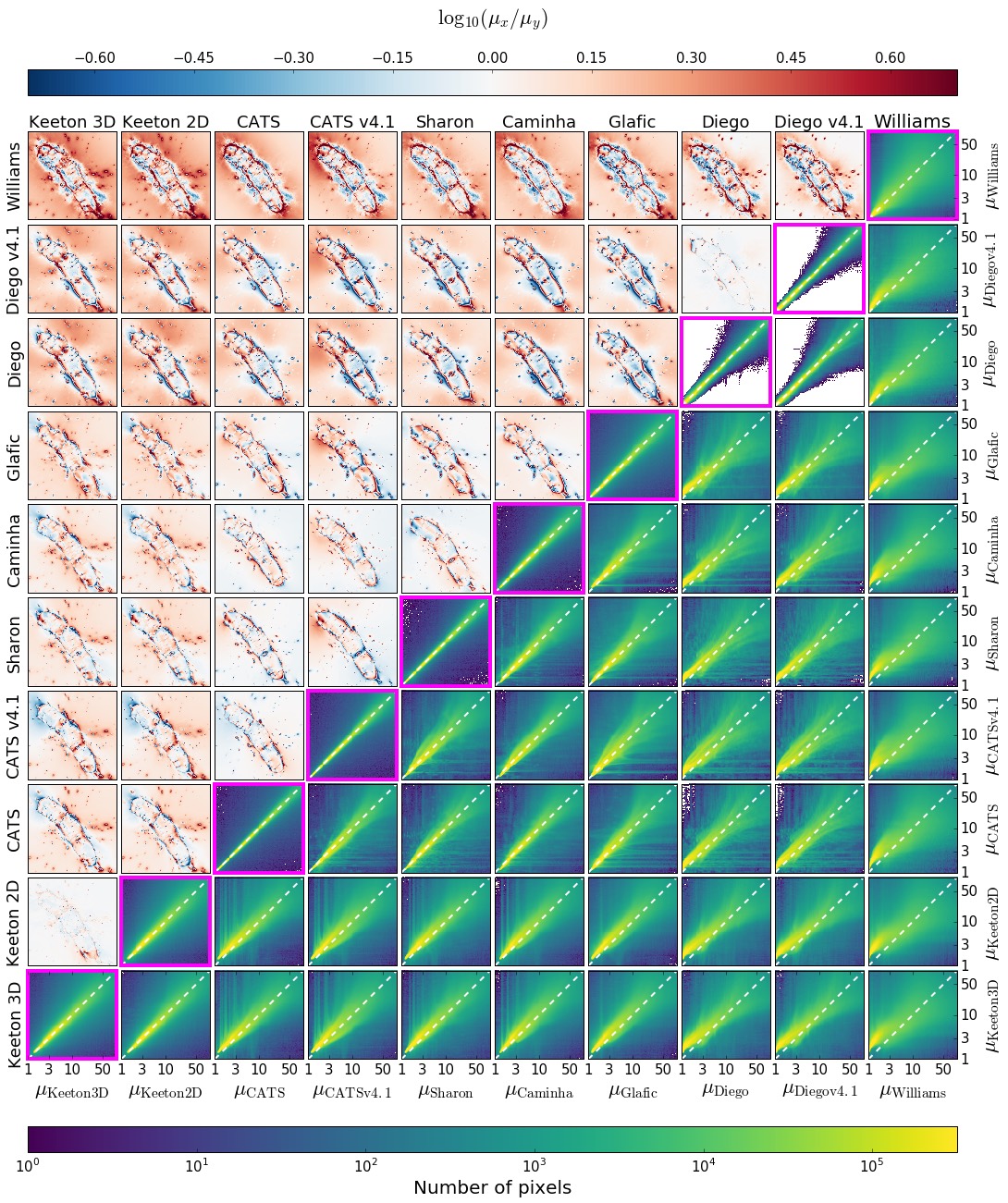}
\caption{Similar to Fig. \ref{fig:a2744-comp} for MACS J0416.}
\label{fig:m0416-comp}
\end{figure*}

Figure \ref{fig:m0416-hsm} shows the HSM maps for the common area among the ten models produced for this field. It is clear that the teams can agree fairly well on where the critical curves sit. This cluster has the highest number of spectroscopically confirmed lensed images out of the six HFF clusters; most models use around 100 images, though Glafic also includes images without spectroscopic redshifts for a total of 202 images. Just as we saw in Abell 2744, the free-form models here have similar structure to the parametric models, specifically in that they find a bend at the northern BCG.

One of the obvious differences in the models comes from their treatment of the cluster members. The number of members included, for example, varies between the teams, as does how mass is assigned to them. For example, galaxies in the Diego models have larger critical curves than the galaxies in the Caminha model. We also see a difference in cluster members between the two CATS models. In Abell 2744, the difference between the CATS v4 and v4.1 models was the rank of constraints used; in MACS J0416, the difference was which galaxies were included in the model. It is clear from Fig. \ref{fig:m0416-hsm} that CATS v4.1 model included galaxies out to a larger radius and indeed, the v4 model has 98 galaxies while v4.1 includes 178. We see the effects of this choice in the ratio panels of Fig. \ref{fig:m0416-comp}. The CATS vs. CATS v4.1 panel shows that there are small differences between these two models, particularly at the northern and southern ends of the cluster, leading to shifts in magnifications. 

In Fig. \ref{fig:m0416-comp} we see that, unlike in Abell 2744, the two Diego models for this cluster agree very well as indicated by the mostly-white ratio panel and the very tight 2-d histogram. Those models also agree more with the parametric models here than they did in Abell 2744. The parametric models here, other than ours, show interesting dipole patterns in their ratio distributions between the northern and southern ends of the cluster. Nonetheless, they overall agree more with each other than with ours or the free-form/hybrid models. 

This is not true when compared to our models, which predict lower magnifications at the northern edge and higher magnifications everywhere else. In \citet{raney2019}, we saw that a model without LOS galaxies was biased low as compared to the 3D model, and here we see that other modeling teams indeed have lower magnifications. This is also borne out in the 2-d histograms. When comparing our models against the other parametric models (bottom two rows), the histograms are populated above the one-to-one (white dashed) line; this is not seen in the other panels comparing parametric models. 

An obvious feature present in the 2-d histograms of Fig. \ref{fig:m0416-comp} is the vertical or horizontal lines in many of the panels. Something similar was seen in Abell 2744, though with thicker lines; it was caused primarily by differences in the position of a large-scale halo outside of the cluster core. The features here are produced by a similar cause, but a different source: galaxy-scale halos. This causes the features to be more numerous since there are more galaxies than large-scale halos, and thinner due to the typical use of scaling relations when assigning mass. 

For example, there are more lines seen in the CATS row than that of the CATS v4.1 due to the former having 80 fewer galaxies. The lines are at different magnifications due to the galaxy's position relative to the cluster's critical curves and thus differing base magnification. Further, the fact that we see this feature only in MACS J0416 and not in other fields, which of course also have galaxies, points to how well the models agree with one another. That is, the features are not getting washed out by differences in the large-scale halos, as they are in the other fields.

\subsection{MACS J0717.5+3745}

\begin{figure*}
\centering
\includegraphics[width=0.7\textwidth]{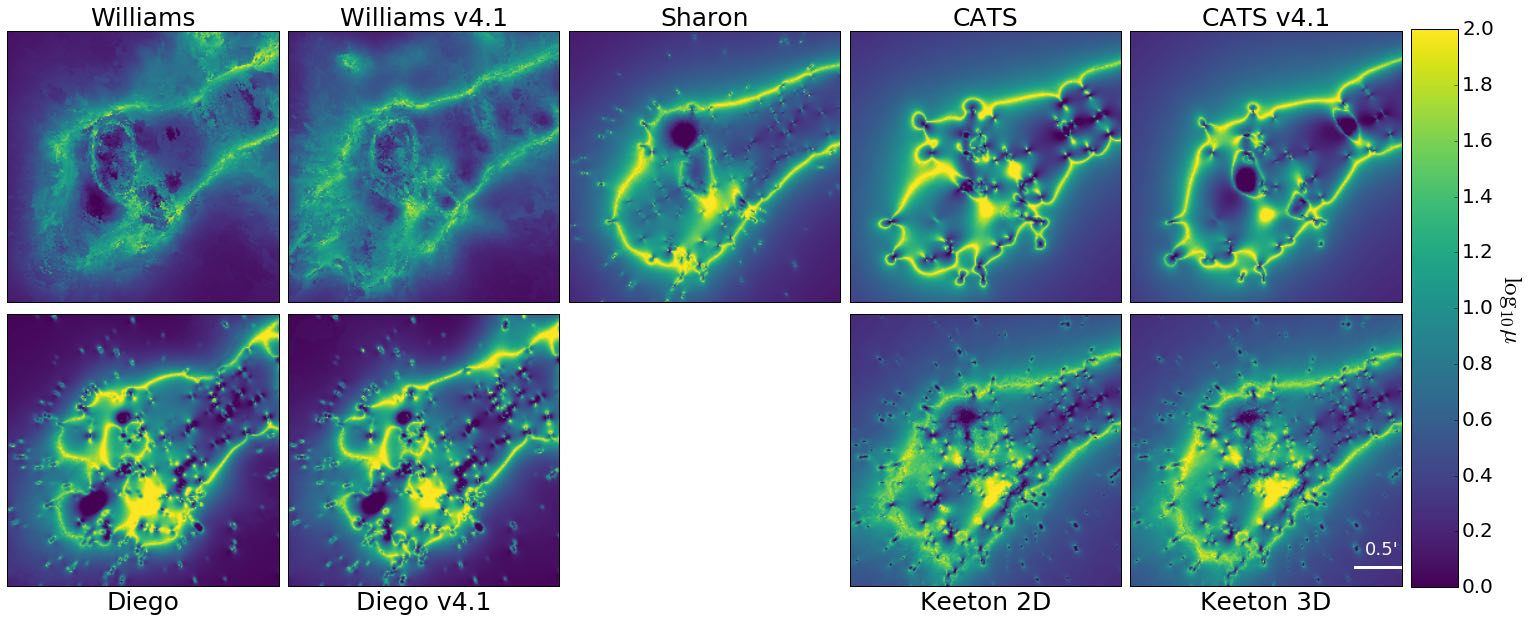}
\caption{Similar to Fig. \ref{fig:a2744-hsm} for MACS J0717.} 
\label{fig:m0717-hsm}
\end{figure*}

\begin{figure*}
\centering
\includegraphics[width=0.9\textwidth]{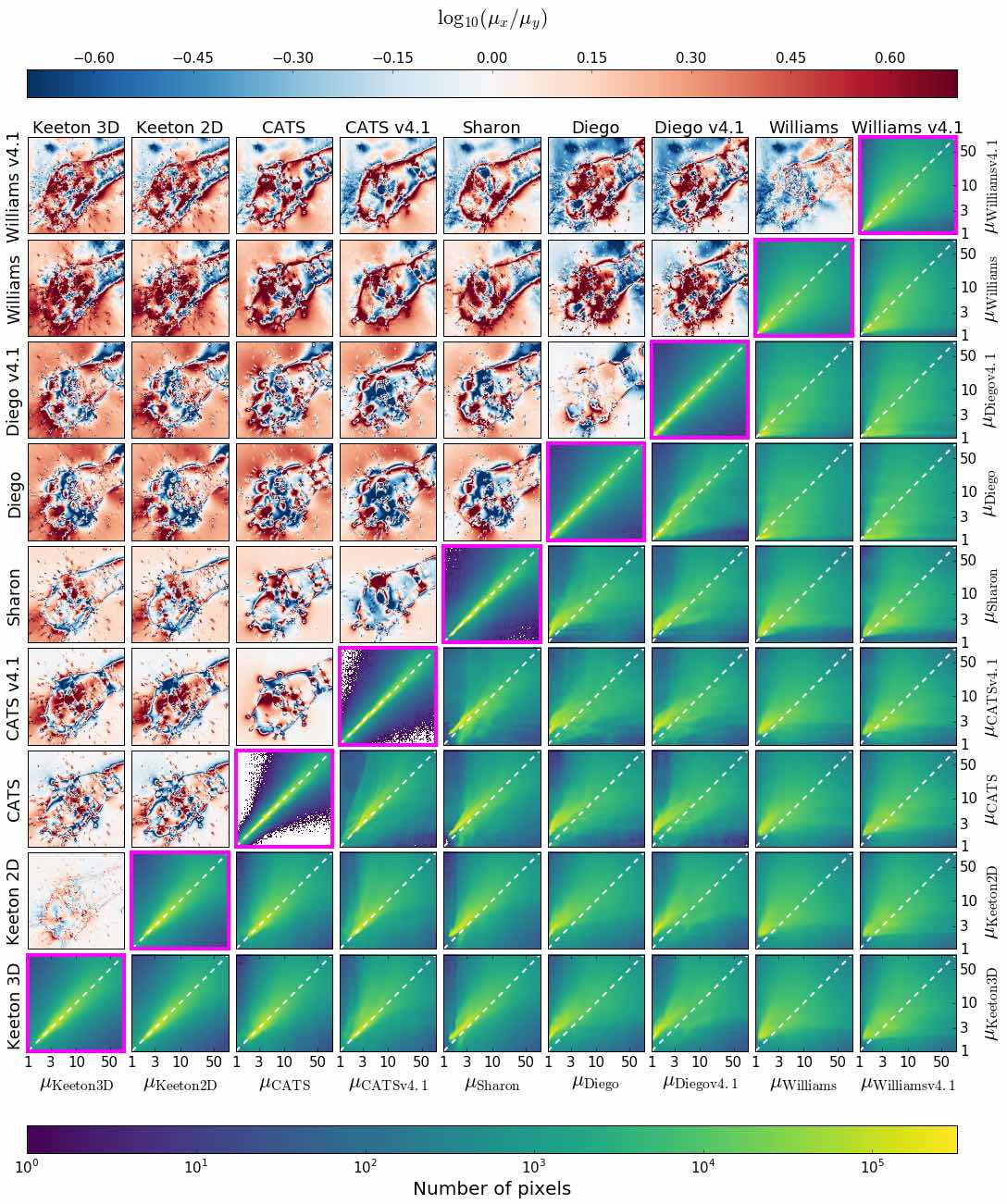}
\caption{Similar to Fig. \ref{fig:a2744-comp} for MACS J0717.}
\label{fig:m0717-comp}
\end{figure*}

This field is very complex, as was seen previously in the disagreements among the modeling teams of the mass profiles. Still, we see an overall structure to the magnification maps that is at least somewhat consistent among the models in Fig. \ref{fig:m0717-hsm}. The critical curves are vaguely mitten-shaped, with all models agreeing on an arm stretching off to the northwest that aligns with the possible filament seen in Fig. \ref{fig:m0717-skymap}.  Contrary to what was seen in the magnification maps of previous clusters, the core of the cluster is not well constrained or agreed upon. This is not surprising given the large disagreement in mass profiles at smaller radii. Indeed, different models show clear offsets between the positions and number of the main halos. All of the models except for those from the CATS team place a massive structure in the middle north of the cluster, though with varying importance. Recall: the CATS models vary from one another in whether the main halos are (v4) or are not (v4.1) cored. 

Another clear difference is seen in the galaxy populations. The CATS team only included the most prominent galaxies in their models, while other teams included more to varying degrees. The size of the critical curves for these galaxies also varies greatly among the models. This could either be due to differing placements of the large-scale halos, or by the varying mass prescriptions used by the teams. The area of low magnification to the southeast of the cluster core in the Diego models is centered on a bright foreground galaxy, which causes further differences in the models.   

The ratio maps, shown in the top left triangle of Fig. \ref{fig:m0717-comp} are expectedly messy near the cluster core. It appears that our models agree more with the cored CATS model than the non-cored v4.1 model, though the Sharon model seems to disagree with both. The two Diego models disagree more with each other in this field than in MACS J0416, but interestingly not as much as in Abell 2744. 

The 2-d histograms of the magnifications in Fig. \ref{fig:m0717-comp} are the broadest of any field, save perhaps for Abell 370. For this field, offsets in halos do not produce clear structures in the panels, e.g. like the ones seen in Abell 2744. This is due to the fact that the halos, though they show clear offsets between teams, are still in the cluster core. We saw the structures in Abell 2744 because the halos of one model fell in a region where the other model did not predict large mass, thus there was a constant small magnification. If both halos are offset but overlapping, this will not be the case, and instead will cause the 2-d histograms to fall in a cloud rather than in nice linear structures. There is also a varying number of halos between each teams that further smears the histograms out.

\subsection{MACS J1149.5+2223}

\begin{figure*}
\centering
\includegraphics[width=0.7\textwidth]{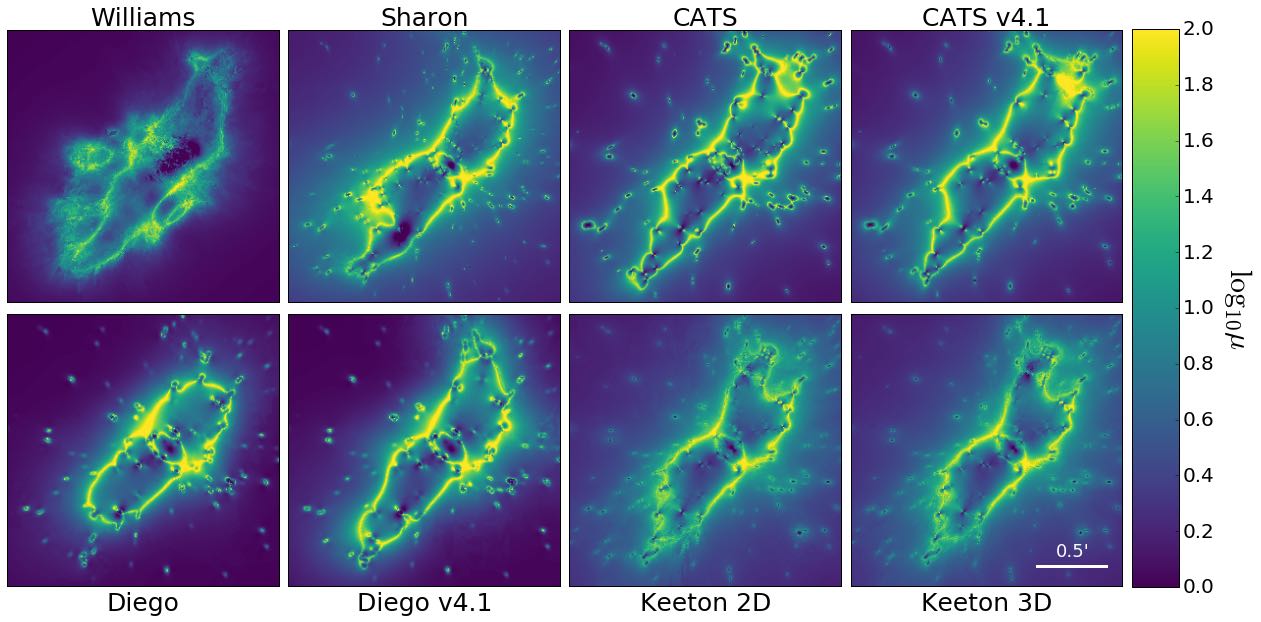}
\caption{Similar to Fig. \ref{fig:a2744-hsm} for MACS J1149. }
\label{fig:m1149-hsm}
\end{figure*}

\begin{figure*}
\centering
\includegraphics[width=0.9\textwidth]{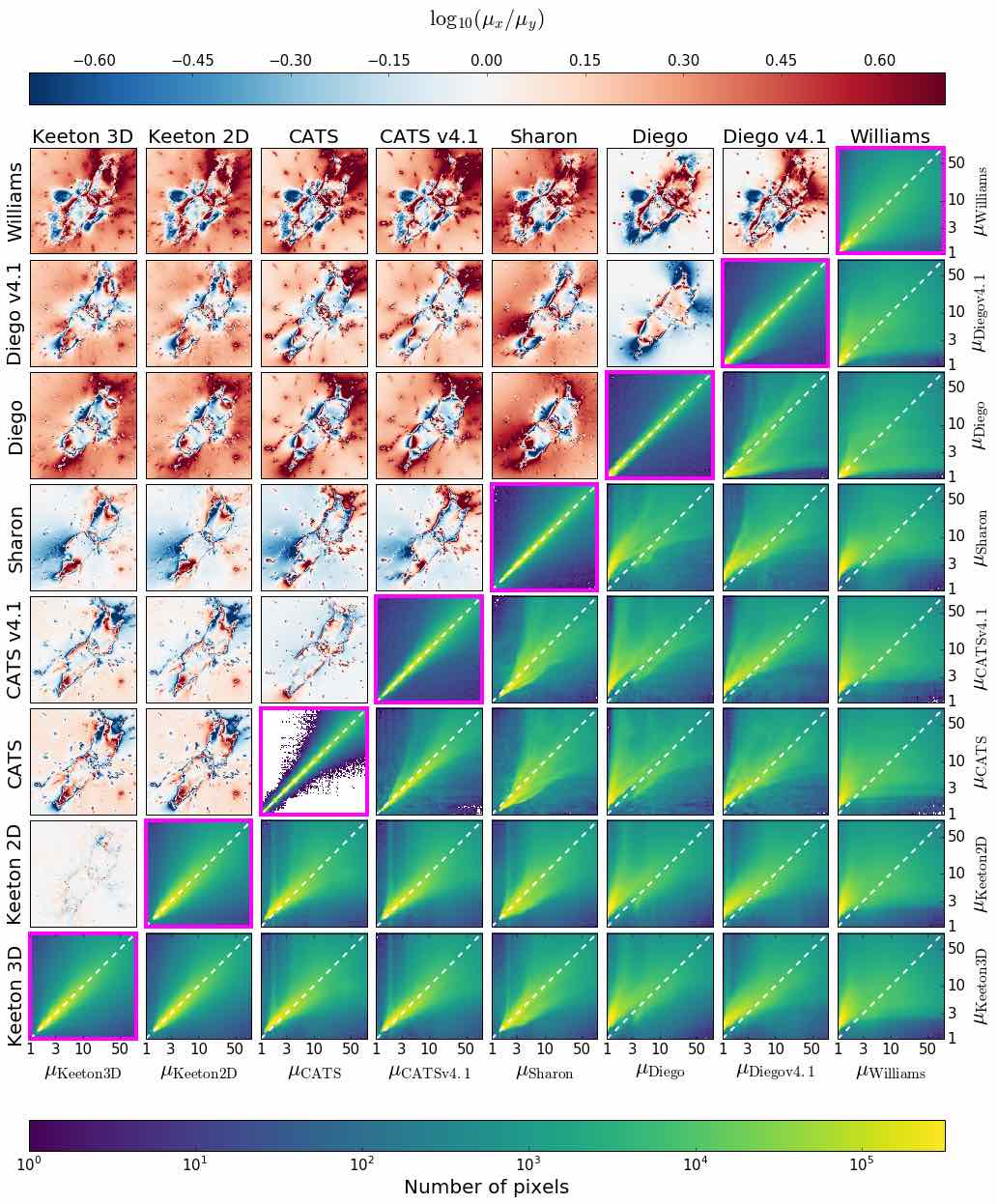}
\caption{Similar to Fig. \ref{fig:a2744-comp} for MACS J1149.}
\label{fig:m1149-comp}
\end{figure*}

This field only had eight models from five teams in the latest round of modeling, likely because the available data did not change much compared to the previous round. Most of the models for this field agree on the broad strokes: the mass distribution is somewhat complicated, with spurs to the north and south off of a vaguely elliptical structure, as shown in Fig. \ref{fig:m1149-hsm}. All models except for Diego v4 agree that this southern region is elongated to some degree, though the Diego v4.1 model shows a more rounded structure than the other models. The Sharon model has a highly concentrated mass component in that area leading to a large area of low magnification. The northern spur is similarly varied, with the Diego model preferring a more rounded structure, while the CATS models have an area of high magnification not seen in the other models. 

In the HSM comparison panels of Fig. \ref{fig:m1149-comp}, we see that magnifications outside the critical curves are essentially one for the free-form models, leading to the red box when comparing the free-form vs. parametric models, as we saw before. The difference in the southern prong between Sharon and the other teams is clear, leading to areas of high magnification ratios. The northern region with high magnification in the CATS models likewise shows a clear divergence from other models, which do not have such an area. 

The locations of the SN Refsdal images are close to the core of the cluster, near the southeastern edge of the ``belt'' of the critical curves. This is in part why the models all agree reasonably well in the middle. It is important to note, however, that those images can only constrain the model at a few points. These models are very complex and can compensate in various ways such that, even if one has images near a dark matter halo at the cluster core to constrain it reasonably well, the models may still disagree on large scales. 

\subsection{Abell S1063}

\begin{figure*}
\centering
\includegraphics[width=0.7\textwidth]{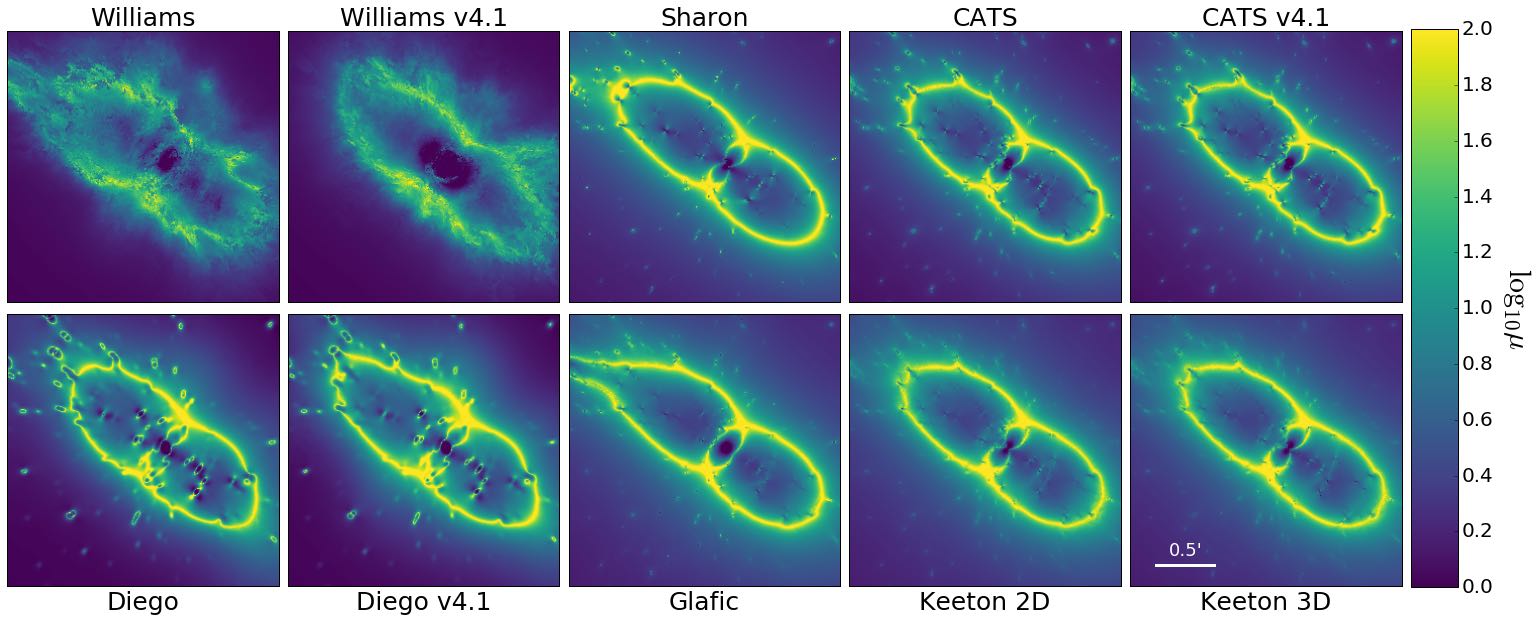}
\caption{Similar to Fig. \ref{fig:a2744-hsm} for Abell S1063. }
\label{fig:a1063-hsm}
\end{figure*}

\begin{figure*}
\centering
\includegraphics[width=0.9\textwidth]{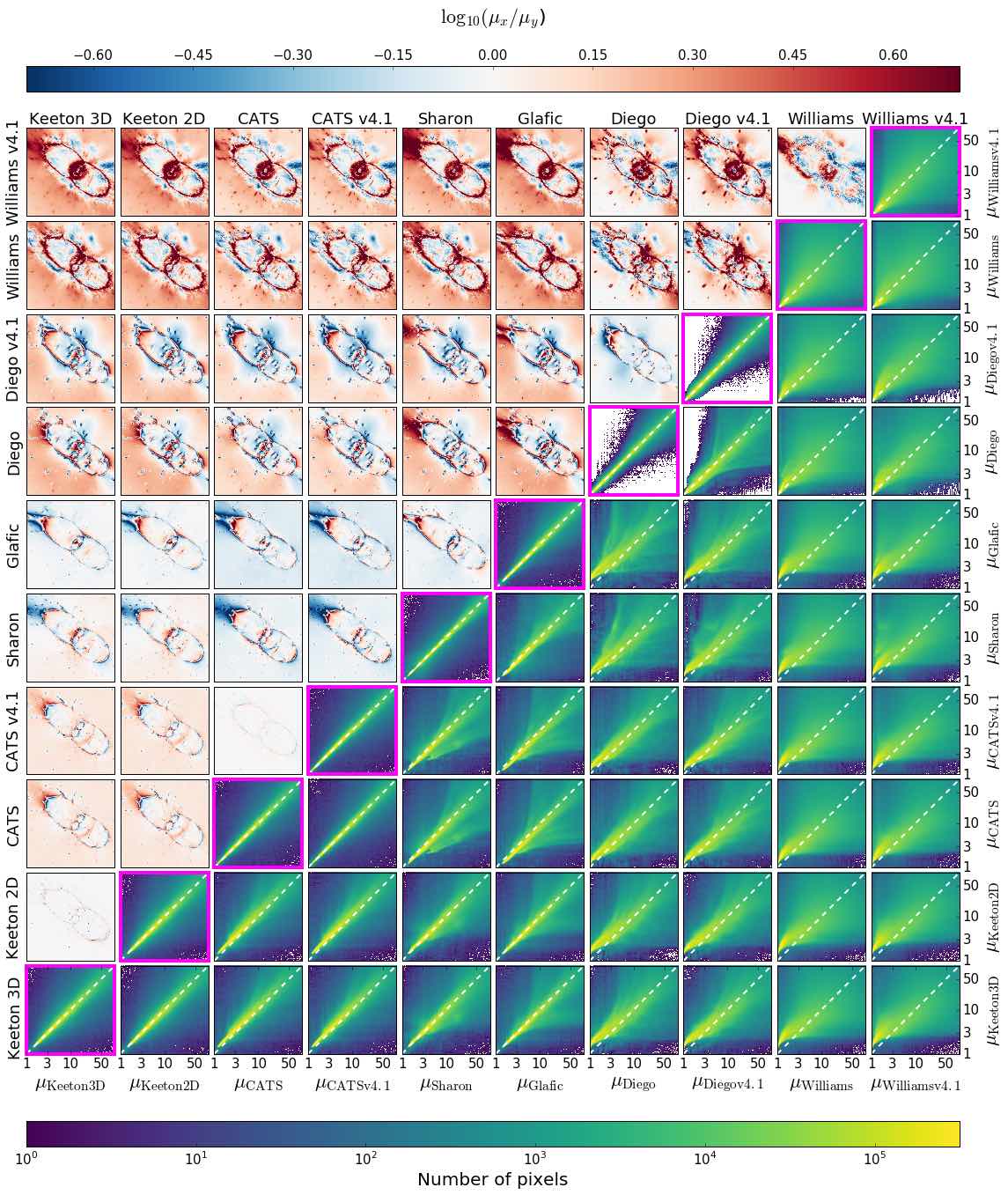}
\caption{Similar to Fig. \ref{fig:a2744-comp} for Abell S1063.}
\label{fig:a1063-comp}
\end{figure*}

This field has very small scatter among the mass profiles, and we see this trend continue into the magnification maps shown in Fig. \ref{fig:a1063-hsm}. Certainly the position angle and ellipticity are well constrained, as is the placement of the ``belt'' at the position of the BCG, even for the free-form models. Of those, the Diego model matches the shape of the parametric models most closely, though with very large critical curves around their galaxies. The Williams v4.1 model has a larger area of low magnification at the core of the cluster than any of the other models. 

The Williams model shows an elongation of the critical curves to the northeast; this horn feature is in the same direction as the elongation seen in the Glafic and, to a somewhat lesser extent, Sharon models. This feature seems to be due to a clustering of member galaxies that are located just out of the bounds of the map; this clustering was also part of the argument by \cite{gomez2012} for a recent merger, thus making it particularly interesting that the models would differ in their treatment of it. 

We note that ours and the two CATS models do not show such an elongation; these models also only have two large-scale halos, whereas at least the Glafic model includes three.  This elongation is further evident in Fig. \ref{fig:a1063-comp}. The Sharon ratio panels show high magnifications compared to all of the other models except for Glafic. Our own models somewhat split the difference between the clustering of galaxies the Sharon, Williams, and Glafic model pick out and galaxies more to the north, similar to, though less drastic than, Diego v4.1. Evidence of this can be seen when comparing our models to the CATS models, which are otherwise very similar in shape. 

The CATS models are interesting in that they have lower magnifications outside of the critical curves than the Sharon models or ours. This is also seen in the 2-d histogram panels of Fig. \ref{fig:a1063-comp} as a shift away from the one-to-one line. The free-form vs. parametric panels exhibit this behavior, as in the other clusters, though in this case the Diego models also appear to be higher than the Williams models. 

\subsection{Abell 370}

\begin{figure*}
\centering
\includegraphics[width=0.7\textwidth]{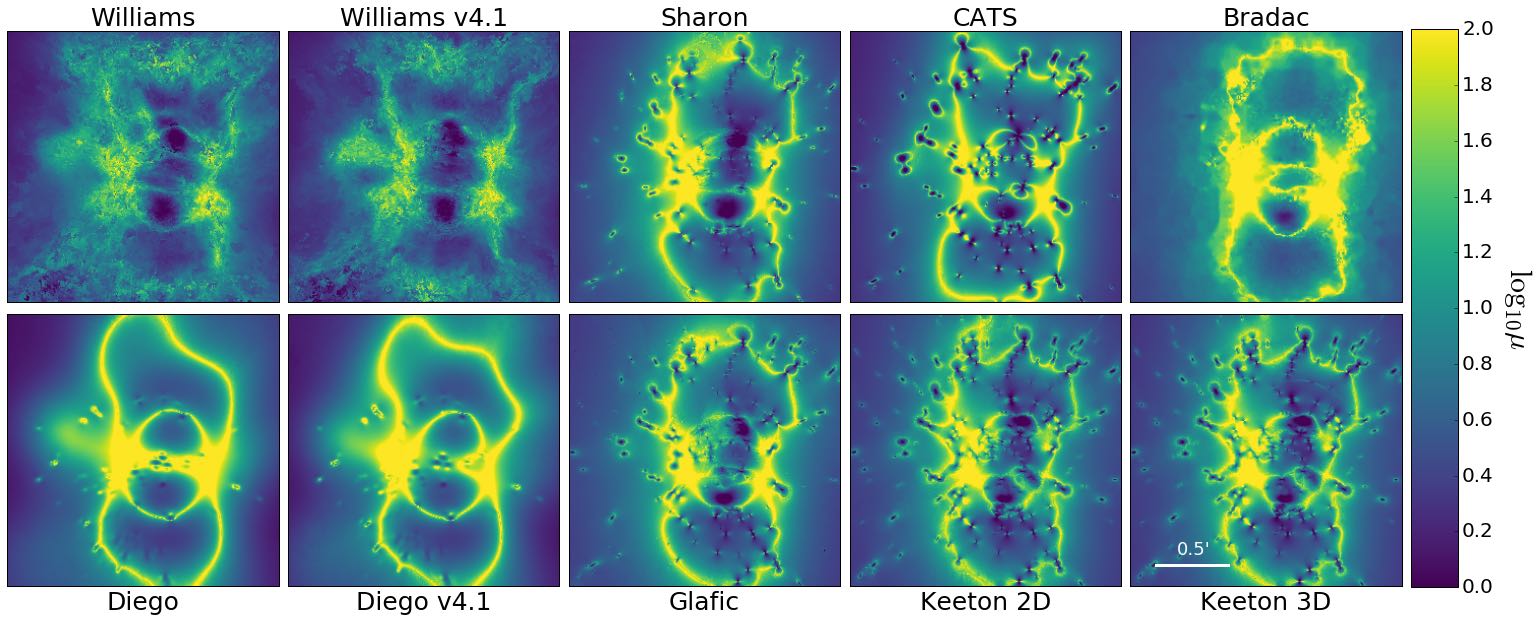}
\caption{Similar to Fig. \ref{fig:a2744-hsm} for Abell 370.}
\label{fig:a370-hsm}
\end{figure*}

\begin{figure*}
\centering
\includegraphics[width=0.9\textwidth]{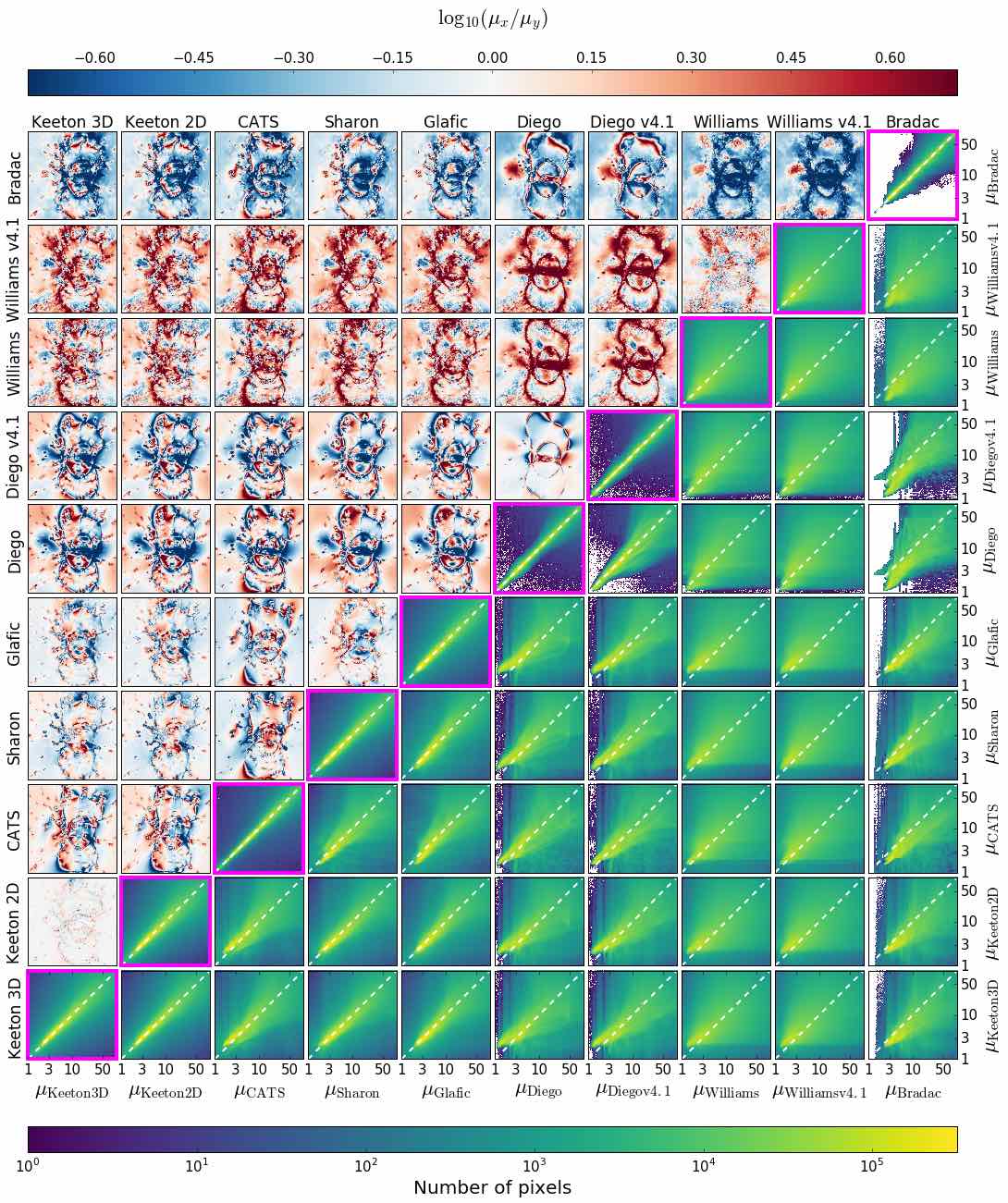}
\caption{Similar to Fig. \ref{fig:a2744-comp} for Abell 370.}
\label{fig:a370-comp}
\end{figure*}

The defining characteristic shared among all of the models of this field is the double-core, as seen in Fig. \ref{fig:a370-hsm}. This duality is caused by the cluster's two BCGs which are very similar in both size and luminosity. This cluster has perhaps the most spread in the Williams models; recall, since the plots shown in Fig. \ref{fig:a370-hsm} are half-sample mode maps, the ``fuzzy'' nature of a plot indicates that there is wide variation in magnifications in that area among the realizations of that model. The models from Diego and Brada\v{c}/Strait do not share this quality, and are tightly constrained, though the Diego models are unique in that they do not have the areas of low magnification near the two BCGs. The Brada\v{c}/Strait model only has low magnifications in the southern lobe, somewhat similar to the CATS model. The Diego model also did not include as many galaxies in their model of this cluster as in other clusters. Our models are different from all others in that they split the southern lobe into two subsections.

The ``crown'' of galaxies in the northern region is asymmetric in the Diego models. The CATS model shows a similar bump caused by the critical curves stretching to a background galaxy with a redshift from GLASS of $z=0.82$ \citep{schmidt2014,treu2015}, as does the Glafic model. We did not include this galaxy due to its distance from images. The Sharon model varies in this region, leading to the fuzzy nature of the HSM map. There is a bright foreground galaxy to the north just out of frame which our 2D model extends up to while our 3D model does not. The Brada\v{c}/Strait model has smoother critical curves in the northern region due to not explicitly including galaxies, though there is a knot in the northeast near a clump of galaxies. 

Our models, along with that of the Sharon team, have a larger high magnification region on the eastern side of the critical curves. This region has such high magnification in our models due to a clustering of galaxies, some of which are foreground galaxies at the same redshift ($z=0.33$), though it is unclear if they are physically related. It is interesting that all four of the Williams and Diego models place a structure extending to the east of the cluster, which is not really seen in the parametric models or the Brada\v{c}/Strait model. This could be a stand in for the cluster members extending off to that side of the cluster, or could perhaps be hinting at some kind of LOS structure that the other parametric models are not taking into account. 

The large area of the high magnifications leads to a wide range in the ratio panels of Fig. \ref{fig:a370-comp}. This is similar to what was seen in MACS J0717, which also had broad swaths of fairly high magnifications. It is important to note that the large differences in the Williams panels are more an artifact of our HSM maps than their modeling process. It is interesting that the Brada\v{c}/Strait model is not part of the red block of the other free-form models which we have seen in every field. It could be due to their different modeling process; recall that they employ weak lensing constraints which would affect the model at large radii. 

 The 2-d histograms offer a similar view of the differences in the models. An interesting characteristic about this cluster is the lack of structure in most of the histograms. This is partially due to the messiness of the cluster, as well as the size, both of which will cause a wide spread in magnifications that leads to a smearing out of the 2-d histograms. This was also seen in MACS J0717, another very messy and large cluster. However, that cluster also had the least number of constraints whereas Abell 370 has the second highest number, just under MACS J0416. Yet the other clusters, barring MACS J0717, have higher agreement between the models. Interestingly, the Brada\v{c}/Strait model has a very tight self-comparison 2-d histogram; in addition, they have virtually no pixels below a certain magnification, leading to a lot of white in their histograms. We see similar behavior in the Diego histograms at low magnifications, though not quite to the same extent. 

There is more spread in the 2-d histograms and structure in the ratio plots for this field than for some of the others. It is clear that this field posed somewhat of a challenge to model, though it is not immediately obvious why. All of the fields in this sample show evidence for a recent or ongoing merger, as evidenced by X-ray studies and/or the fact that they have more than one BCG; Abell 370 is certainly not unique in this regard. However, it is notable that this mass distribution is physically wider than the other fields. For example, Abell 2744, MACS J0416, and MACS J1149 are all fairly thin on the short axis. MACS J0416, which has two BCGs just as Abell 370 does and about as many lensed images, is about an arcminute on its short axis; Abell 370 is around twice that.

\section{Discussion}
\begin{figure*}
\centering
\includegraphics[width=0.9\textwidth]{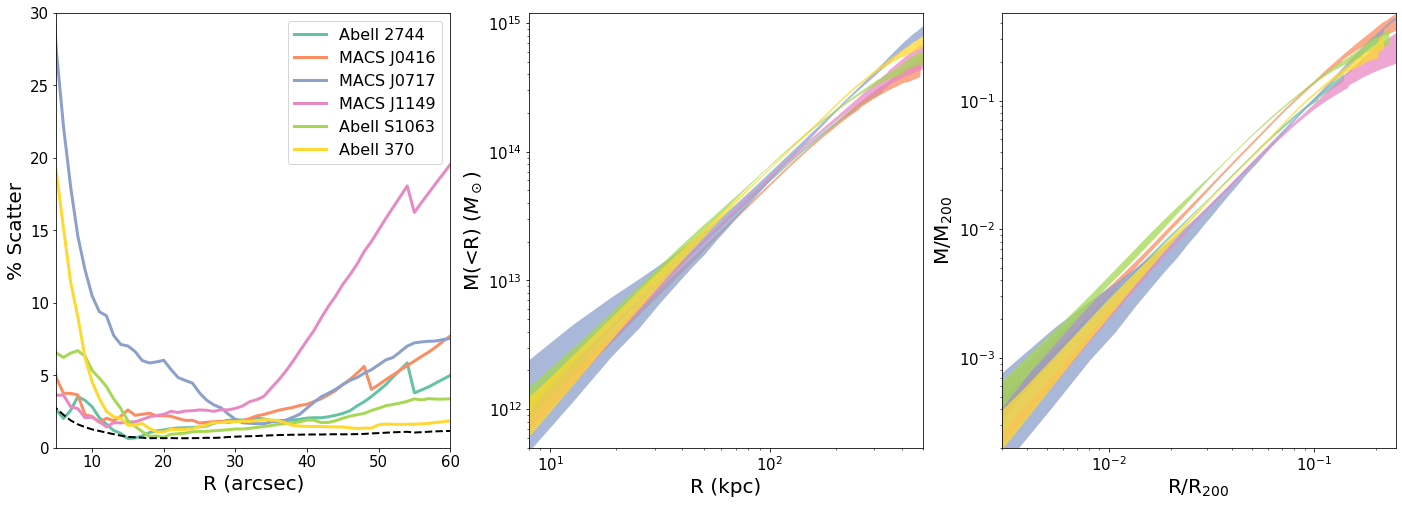}
\caption{\textit{Left:} Scatter given by the $1\sigma$ error across all realizations of the submitted models for each field, as a function of radius in arcseconds. The black dashed line is the average statistical scatter across all models and all fields. \textit{Middle:} Median enclosed mass (given in solar masses) as a function of physical radius in kiloparsecs. The $1\sigma$ error bars are taken across all submitted models for a given field. Colors are the same as in the left panel. \textit{Right:} Similar to the left panel, but now scaled by $M_{200c}$ and $R_{200c}$ values. These were obtained from \citet{medezinski2016} for Abell 2744, \citet{umetsu2011} for Abell 370, and \citet{umetsu2016} for the rest.}
\label{fig:mass-dist}
\end{figure*}

\begin{figure*}
\centering
\includegraphics[width=0.9\textwidth]{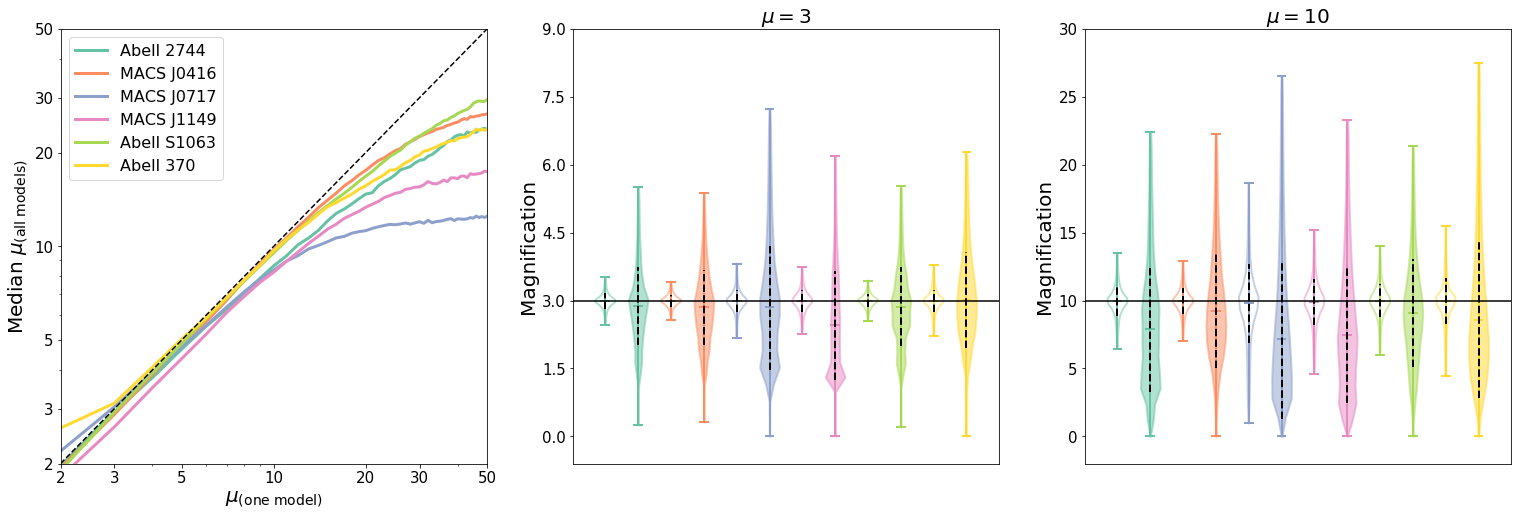}
\caption{\textit{Left:} For a given magnification, we show how well the magnification predicted by one model tracks the magnification predicted by all models. Specifically, we look at the pixels from one model that have a given magnification ($\mu_{(\mathrm{one\;model})}$) and then find the distribution of magnifications for those pixels across all realizations of the other models, i.e.  $\mu_{(\mathrm{all\;model})}$. The plotted curve is then the median of this distribution. If a field has many models with similar magnifications, then its curve will fall close to the one-to-one line (black, dashed). \textit{Middle:} We show the full distribution of $\mu_{(\mathrm{all\;model})}$ for a magnification of 3. The median values will be the same as the left panel, but now the full distribution is shown as well. Unfilled violin bodies are when the magnifications of one model are compared against itself, analogous to the panels along the diagonal in the 2-d histograms. Dashed black lines show the 1$\sigma$ error. Colors are the same as in the left panel. \textit{Right:} Similar to the middle panel, but for a magnification of 10.}
\label{fig:mag-dist}
\end{figure*}

\subsection{Mass Profiles}

One of the ways in which strong lensing can be a tool is in determining how mass is distributed within a system. It is an interesting exercise to see how well lens models can reproduce distributions that are complicated: for example, a cluster in a state of merging. Further, it is important to see how the results differ between modeling teams who use different techniques and density profiles to assign mass to their halos. \citet{meneghetti2017} studied this by creating two mock clusters and asking numerous teams to model them as a way to perform a controlled test. The fields were created using two different methods, though it is important to note that both used the light-traces-mass assumption, which is also often employed by parametric modeling methods. They found that the teams using parametric methods were able to reproduce the true mass profiles to within $\pm2-10$\%, while the free-form teams had slightly higher scatter of $\pm5-15$\%. This was true even though some teams did not use the same density profile as the input mock model did. 

In Sec. \ref{sec:mass-comp}, we showed the mass profiles for submitted models of each field. From the plots, it was clear that the models were, for the most part, well constrained and showed little scatter between the models. We quantify this in the left panel of Fig. \ref{fig:mass-dist} by showing the percent scatter across all the realizations for each field out to an arcminute from the BCG. This is found by taking the half-width of the 68\% confidence interval across all realizations and dividing by the median. The features seen in the mass profiles are also borne out here. For example, MACS J0717 clearly has the largest scatter at low radii, partially due to the cored vs. non-cored models of the CATS team, which fit the data equally well. Abell 370 also showed high scatter at low radii, but it quickly falls off to the lowest values of all six fields. At larger radii, MACS J1149 has a scatter that is more than twice the other fields, likely because the area spanned by the lensed images is the smallest of the sample.

Nonetheless, we find that the scatter is quite often below 5\%, which is somewhat remarkable given that these clusters are very complicated and often in various stages of merging. In the left panel of Fig. \ref{fig:mass-dist}, we also show the mean statistical error for all six fields in the black dashed line, found by averaging error in a given model using the realizations. Though some of the other curves get quite close to this line, most are indeed above it. This suggests that systematics between the models are more important than statistical uncertainty, which has been a known problem in cluster lensing and which we will again see among the magnifications.

With these models, we can also ask how the mass profiles of the clusters compare to one another. In the middle panel of Fig. \ref{fig:mass-dist}, we show median mass profiles (across all submitted models) now as a function of physical radius in kiloparsecs, along with $1\sigma$ error bars. The error bar is quite large at low radii for MACS J0717, which is unsurprising given the left panel. However, as radius increases and thus more lensed images are included within the radius, the error shrinks. Across all of the six fields, at 100 (200) kpc from the BCG, the mean enclosed mass is $0.668 \times 10^{14} M_{\odot} \pm 11\%$ ($1.96\times 10^{14} M_{\odot} \pm 12\%$). 

Past studies of simulations \citep{diemer2014} have shown that clusters should be self-similar, and thus should also have very similar mass profiles, specifically when scaled by $M_{200c}$ and $R_{200c}$. Indeed, a recent study by  \cite{caminha2019} examined clusters from the Cluster Lensing And Supernova survey with Hubble \citep[CLASH;][]{postman2012} and found just that: among profiles of seven clusters, the scatter was only 5-6\%. 

We sought to test this with our own profiles, as shown in the right panel of Fig. \ref{fig:mass-dist}. While the two clusters included in \cite{caminha2019} are quite similar (MACS J0416 and Abell S1063), the others are fairly different. This causes a slight decrease in the average enclosed mass we find as compared to values reported in \cite{caminha2019}. We also find an increase in scatter: around 15\%. Interestingly, the scatter is slightly larger in this case as opposed to the unscaled case. We note that this does not include the error in the $M_{200c}$ or $R_{200c}$ measurements, which can be $\sim25\%$.

\subsection{Magnification Maps}
Among the magnifications, we often do not find the remarkable similarity seen in the mass profiles. Since the goal of the HFF program was to find high redshift galaxies, understanding magnification errors is vital given that these errors may propagate into luminosity function calculations. Multiple teams were invited to model the fields so that the error in magnification could be estimated by considering the various models. It is important to note that most cluster lenses do not have the same modeling effort behind them. We can then use the HFF models to ask how we might be biasing our magnification estimates by using only one lens model of a given field. 

Essentially we want to find the conditional probability $P(\mu\,|\,\mu_{ref})$ of finding a magnification $\mu$ across all models given that one model predicts a magnification of $\mu_{ref}$. For this analysis, we take a given realization of a model as our reference and find all pixels in that map that have a certain magnification, say $\mu_{ref}=3$. We then look at magnifications for that set of pixels across all realizations of the other models. We can repeat this procedure, changing which model and realization we use as our reference, creating a distribution of magnifications. If the models all agree with each other, i.e. if one model has high predictive power for the other models, then the distribution should be tightly constrained around $\mu_{ref}$. 

In the left panel of Fig. \ref{fig:mag-dist}, we show the median of this distribution across all models of the six fields versus the reference magnification from any one given model. The black, dashed line is one-to-one and illustrates magnifications from one model perfectly agreeing with median magnifications across the other models. We find that, at low magnifications, one model can predict the median magnification fairly well. However, it does start diverging at higher magnifications. Different fields are affected at different times: e.g., Abell 2744, MACS J0717, and MACS J1149 are farther away from the one-on-one line at $\mu=10$ than the other three fields. At large magnifications, the difference is large for all six fields. 

We note that the curves in the left panel of Fig. \ref{fig:mag-dist} are mostly below the one-to-one line, suggesting that, at a given pixel with a high magnification in one model, the other models will predict a lower magnification. This has to do with the non-linear nature of magnification and, specifically, the critical curves. Since magnification drops off quickly as one moves away from a critical curve, you have many more low magnification pixels than high magnification, which causes this bias towards lower magnifications. The effect grows with magnification as well, which causes the flattening of the curves. We explore this further using a toy model in Appendix A. 

In the middle and right panels of Fig. \ref{fig:mag-dist}, we use violin plots to depict the full distribution of magnifications for reference values of 3 and 10. For comparison, we also isolate the statistical scatter via the unfilled violin plots. That is, we now look at all pixels where $\mu=3$ or 10 for a model and consider the distribution that consists of the magnifications at those pixels across only the realizations of that model, as opposed to the realizations across all models (which are shown in the filled violin plots). 

As we saw in the left panel, the medians are further away from the correct value at $\mu=10$ than $\mu=3$. One can also see that the scatter is much larger in the higher magnification case. \citet{priewe2017} did a similar analysis of the results for two fields, Abell 2744 and MACS J0416, from the v3 round of modeling. They found a scatter of 30\% at low magnifications ($\mu\sim2$), which increased to 70\% at higher magnifications ($\mu\sim40$). We find a similar amount of scatter for these fields, along with Abell S1063, for our low magnification case of $\mu=3$. Abell 370 has a slightly higher amount of scatter at 35\%, but the largest scatter lies in our highest redshift clusters, MACS J0717 and MACS J1149, which both show 49\% scatter at low magnification. We also note that the average statistical scatter across all six clusters is significantly lower at $\sim 6\%$.

For the higher magnification case, $\mu=10$, the amount of scatter is, unsurprisingly, even higher. The lowest scatter is seen in MACS J0416 and Abell S1063 at $\sim45\%$, while the highest is in MACS J0717: 82\%. The other three clusters range from $59\sim67\%$. Statistical scatter is still far below these values, though it does increase: for $\mu=10$, $\sigma_{syst}=4.1\times\sigma_{stat}$, as opposed to $5.7\times\sigma_{stat}$ for the $\mu=3$ case. 

It is interesting that our results agree with those of \citet{priewe2017}, given that there were significant changes to the constraints of Abell 2744 and MACS J0416 between v3 and v4. Specifically, two surveys utilizing VLT/MUSE greatly increased the number of spectroscopic constraints for the fields. In Abell 2744, the number of image families with spectroscopic redshifts went from 5 to 29 \citep{mahler2018}, and from 15 to 37 in MACS J0416 \citep{caminha2017}.

However, this does seem to be in line with the work of \citet{johnson2016}, which considered how the number and type of constraints impacted model fits for the two mock clusters presented in \citet{meneghetti2017}. They found that there was a limit to how much additional constraints decreased magnification error in the models; specifically, the decreasing error leveled off around 25 image systems. Further, magnification bias or variance did not correlate with fraction of images with spectroscopic redshifts as long as the constraints included at least five spectroscopically-confirmed systems. Those models without any spectroscopic constraints had magnifications biased low; this could be explained by an increase in model variation, which we have previously shown will decrease magnifications. They also found that exactly which image systems are used as constraints can be a bigger source of systematic error than number of spectroscopic redshifts. This could be a important part of the systematic error we see here, given that there is such a wide range in constraint selection between the teams.

Other works have looked at how these errors propagate into luminosity functions, finding various results. For instance, \citet{livermore2017} found that magnification uncertainties did not have a large effect on the luminosity function, while \citet{bouwens2017} found that a large uncertainty could produce an artificial steepening of the slope. \citet{atek2018} used the submitted models of each team to get error bars on their measurements, though we note that this technique would not be possible if there were not multiple modeling teams. 

\subsection{LOS effects}
In \cite{raney2019}, we showed that there can be systematic effects produced when galaxies along the line-of-sight to a cluster are either not included in the model or their effects are approximated to the cluster lens plane. Specifically, not placing the galaxies at their true redshift could cause a bias in magnifications on the level of 5\% or could cause an increase in the scatter of the magnifications. We argued that, while these effects were non-negligible, they were also quite small and unlikely to be the dominant source of error in current models.

In this magnification analysis of this work, we included both the model where galaxies are approximated to the cluster lens plane (Keeton 2D) and the case where these LOS galaxies are placed at their true redshift and thus the model has multiple lens planes (Keeton 3D). We find that the results from our previous work are again supported here. Some small differences can be seen between these two models. For example, in the HSM ratio Keeton 2D vs. 3D panel of MACS J0416 (see Fig. \ref{fig:m0416-comp}), there is a knot in the southern part of the cluster where magnifications are quite different that coincides with the location of a bright foreground galaxy. Another example can be seen in the HSM ratio Keeton 2D vs. 3D panel of Abell 2744 (see Fig. \ref{fig:a2744-comp}) where there is a very slight blue tinge across the plot; this corresponds to the 2D model predicting lower magnifications than the 3D model, as we saw in our previous work.

Abell 2744 is an interesting case as well because there is actually more of a difference between the Keeton 2D and 3D models than there is between the CATS v4 and v4.1 models. Recall, the difference between the two CATS models is their lensing constraints. This could again support the previous assertion that adding additional constraints does not significantly change the magnifications of the model. If we are indeed in the regime where additional constraints are not useful in further constraining models, it is then worrying that the different models among the teams are not more similar. This could be a problem for future cluster lensing surveys, which will likely not have the same modeling effort the HFF did. We note that this seems to only be true of the parametric models of Abell 2744: the two Diego models, which also vary in constraints used, do show many differences in their magnification maps. 

\section{Conclusions and Implications} \label{sec:results}

The HFF program was a tremendous effort by many. It took a significant amount of observing time, with both HST and other telescopes, in order to conduct the photometric and spectroscopic surveys needed. Also, the different lensing teams put in the effort to find and rank the possible lensed images and, of course, model the fields. Thus it serves as a wonderful opportunity to compare the results of the models of these fields and see what the state of the lensing field is. Though the HFF program has finished, further cluster lensing surveys are underway: the Reionization Lensing Cluster Survey \citep[RELICS;][]{coe2019} and Beyond Ultra-deep Frontier Fields and Legacy Observations \citep[BUFFALO;][]{steinhardt2020}, the successor to HFF. We can then use the results from the HFF modeling effort to make improvements going forward.

We chose to compare the models in two ways in this work: circularly-averaged mass profiles, derived from the surface density maps, and magnifications. These models came from eight teams using a variety of different methodologies and making various decisions in the modeling process. We considered not just the fiducial models, but also the realizations that each team submitted. In this way, we were able to get an idea of how systematic errors compare to the statistical errors of each team. The conclusions we drew can be summarized as follows:
\begin{itemize}
\item The circularly-averaged mass profiles are remarkably similar across the models with $1\sigma$ scatter often $<5\%$. This systematic scatter across all models is larger than the statistical error for a given model, though in some cases it is quite close.
\item The mass profiles across fields are also notably similar to one another when plotted as a function of physical radius in kiloparsecs, with a scatter of only about 13\%. They become less similar when scaled by M$_{200c}$ and $R_{200c}$, and the scatter becomes 20\%. 
\item Magnification maps often show significant differences between teams. If one assumes a single model is correct and compares magnifications at a given pixel, results will be biased low due to the non-linear nature of magnifications maps. This bias is fairly small at low magnifications, where the median magnification averaged across the six fields is 2.82 for $\mu=3$. However, the bias increases with magnification: $\mu=10$ gives a median magnification of 8.22.
\item Further, the scatter in these magnifications can be quite high: $30\sim50\%$ at low magnifications and $45\sim82\%$ at higher magnifications. This large uncertainty may propagate into quantities derived using magnification, i.e. intrinsic luminosity or size of the lensed galaxies.
\end{itemize}

Is it worrying that, even with dozens to hundreds of lensed images per field, the models still show clear disagreements? It certainly suggests that statistical uncertainties have decreased to the point that they are smaller than systematic effects in lens modeling. This is an important lesson because it is not likely that future surveys will have 5+ lens modeling teams to sample the systematic effects. We need to use this opportunity provided by the HFF program to thoroughly understand the systematics in cluster lens modeling and ensure that uncertainties in future surveys are not underestimated.

At the same time, we believe it is still impressive that such complicated systems can be modeled with the precision seen. Perhaps another lesson involves the choice of systems for detailed study. The models of Abell S1063, a fairly simple cluster, show the most agreement among the teams. While larger clusters such as MACS J0717 may have larger areas of high magnification, they are much harder to study and to constrain the lens models, leading to higher error bars on the luminosities of any high redshift galaxies found. It is an interesting question for the future of cluster lensing: should we focus more on those fields which are large and massive (thus very likely to have elongated areas of high magnification) even though they also may be very complicated due to mergers? Or, instead, would it be better to look at neater fields that are easier to model, even if they lack the lensing power seen in the more complicated systems? 

There is much work that could be done in the future. Particularly, there are many sources of systematic error that have not been studied in great detail. Further, which systematic biases are most important (and the strength of those biases, as we showed in our previous work) may depend on the particular cluster. Thus any study of systematics should ideally be done for more than one or two fields. The next generation of telescopes will be promising for cluster lensing, e.g. \textit{JWST} and \textit{WFIRST} with their IR capabilities to find high redshift galaxies and, in the latter case, a wide field-of-view to study mass in the cluster outskirts. With more and more data, work into quantifying systematic errors will become vital if we are to use these fields to their full potential as ways to detect and study galaxies from the early Universe. 

\section*{Acknowledgements}
We recognize the other Hubble Frontier Fields version 4 modeling teams whose work was used in this paper: Bradac \& Strait; Caminha; Richard, Natarajan \& Kneib (CATS); Diego; Ishigaki, Kawamata, and Oguri (Glafic); Sharon \& Johnson; and Williams. We acknowledge support from Hubble Frontier Field Lensing Support through contract STScI-49745 from the Space Telescope Science Institute, which is operated by NASA under Contract No. NAS5-26555. CAR and CRK also acknowledge support from the US National Science Foundation through grant AST-1909217. All version models were obtained via the Mikulski Archive for Space Telescopes (MAST) and the web-based lens model tool. We thank Dan Coe and Keren Sharon for organizing the useful discussions among all modeling teams and contributors. We thank John Hughes and Peter Doze for their helpful discussions about cluster mass profiles. We acknowledge Nicholas James for his work on the 2-d histograms which served as an initial step for some of the plots shown in this paper. 



\bibliographystyle{mnras}
\bibliography{bib} % if your bibtex file is called example.bib

%%%%%%%%%%%%%%%%%%%%%%%%%%%%%%%%%%%%%%%%%%%%%%%%%%

%%%%%%%%%%%%%%%%% APPENDICES %%%%%%%%%%%%%%%%%%%%%
\appendix
\section{Magnification Distributions with Scatter}\label{sec:mag-dist}
\begin{figure}
\centering
\includegraphics[width=0.51\textwidth]{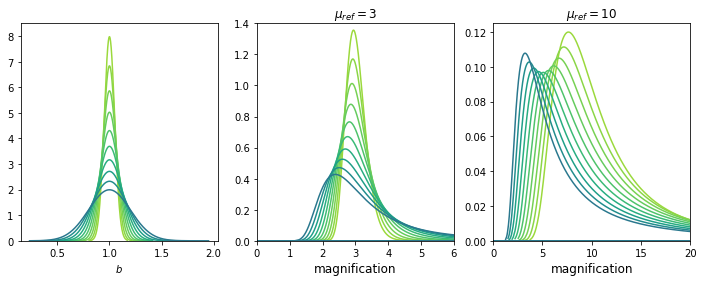}
\caption{\textit{Left:} Probability distribution function for the Einstein radius $b$ with standard deviation $\sigma_b=[0.05,0.2]$. \textit{Center}: Magnification distribution for a reference magnification of three. As $\sigma_b$ increase, the distribution shifts towards smaller values and becomes wider. \textit{Right:} Magnification distribution for a reference magnification of 10. We see behavior similar to the middle panel, but more dramatic.}
\label{fig:appendix}
\end{figure}

The result in Fig. \ref{fig:mag-dist} is perhaps counterintuitive: regardless of which model is chosen as the reference, all other models tend to predict a lower magnification at the reference pixels. To explain this, we consider a toy model of an isothermal sphere. In general, the magnification for an SIS with Einstein radius $b$ is $\mu(r)$ such that
\begin{equation}
\mu^{-1} = 1-\frac{b}{r}.
\end{equation}
Let the reference model have Einstein radius $b_0$, and consider the radius where the magnification is $\mu_0$. Then for another model with Einstein radius $b$, the magnification at that same radius is
\begin{equation}
\mu^{-1} = 1-\frac{b}{b_0}(1-\mu_0^{-1}).
\end{equation}

Now let $b$ be drawn from a Gaussian distribution with varying standard deviation $\sigma_b$, as shown in Fig. \ref{fig:appendix}. The resulting magnification distributions are shown for two reference magnifications, 3 and 10, in the middle and right panels respectively. As the error in the Einstein radius increases, the magnifications shift towards smaller values. Further, the effect is stronger for the higher magnification case, as was seen in Fig. \ref{fig:mag-dist}.

For the models of the HFF clusters, there are multiple parameters that have varying error associated with them, not just the Einstein radius parameter. However, this toy model with one source of error still provides a valuable result. Namely, the increase in the error of the parameter disproportionately affects higher magnifications: the highest $\sigma_b$ (0.20) results in a shift of the median for the $\mu_{ref}=10$ case to $\mu=4.1$, while the lower magnification case still has a median magnification of $\mu=3$. 

%%%%%%%%%%%%%%%%%%%%%%%%%%%%%%%%%%%%%%%%%%%%%%%%%%

% Don't change these lines
\bsp	% typesetting comment
\label{lastpage}
\end{document}